\documentclass[letterpaper,twocolumn,10pt]{article}
\usepackage{usenix-2020-09}

\usepackage{tikz}
\usepackage{amsmath}
\usepackage{graphicx}
\usepackage{wrapfig}
\usepackage{tabularray}
\usepackage[flushleft]{threeparttable}
\usepackage{subcaption,booktabs}
\usepackage{float} 
\usepackage{stfloats} 
\usepackage{shortcuts}
\usepackage{makecell}
\usepackage{multirow}
\usepackage{algorithm}
\usepackage{algpseudocode}

\begin{document}

\date{}

\title{Client-side Gradient Inversion Against Federated Learning from Poisoning}

\author{ 
{\rm Jiaheng Wei} \\
{\rm RMIT University} \\
\and
{\rm YanJun Zhang} \\
{\rm University of Technology Sydney} \\
\and
{\rm Leo Yu Zhang} \\
{\rm Griffith University} \\
\and
{\rm Chao Chen} \\
{\rm RMIT University} \\
\and
{\rm Shirui Pan} \\
{\rm Griffith University} \\
\and
{\rm Kok-Leong Ong} \\
{\rm RMIT University} \\
\and
{\rm Jun Zhang} \\
{\rm Swinburne University of Technology} \\
\and
{\rm Yang Xiang} \\
{\rm Swinburne University of Technology} \\
}

\maketitle

\begin{abstract}
Federated Learning (FL) enables distributed participants (e.g., mobile devices) to train a global model without sharing data directly to a central server. Recent studies have revealed that FL is vulnerable to gradient inversion attack (GIA), which aims to reconstruct the original training samples and poses high risk against the privacy of clients in FL. 
However, most existing GIAs necessitate control over the server and rely on strong prior knowledge including batch normalization and data distribution information. 
In this work, we propose \textbf{C}lient-side poisoning \textbf{G}radient \textbf{I}nversion (\attack), which is a novel attack method that can be launched from clients. 
For the first time, we show the feasibility of a client-side adversary with limited knowledge being able to recover the training samples from the aggregated global model. 
We take a distinct approach in which the adversary utilizes a malicious model that amplifies the loss of a specific targeted class of interest. 
When honest clients employ the poisoned global model, the gradients of samples belonging to the targeted class are magnified, making them the dominant factor in the aggregated update. 
This enables the adversary to effectively reconstruct the private input belonging to other clients using the aggregated update.
In addition, our \attack also features its ability to remain
stealthy against Byzantine-robust aggregation rules (AGRs). 
By optimizing malicious updates and blending benign updates with a malicious replacement vector, our method remains undetected by these defense mechanisms. 
To evaluate the performance of \attack, we conduct experiments on various benchmark datasets, considering representative Byzantine-robust AGRs, and exploring diverse FL settings with different levels of adversary knowledge about the data. Our results demonstrate that \attack consistently and successfully extracts training input in all tested scenarios. 

\end{abstract}

\section{Introduction}
Federated Learning (FL)~\cite{mcmahan_communication-efficient_2017,konecny_federated_2016,shokri_privacy-preserving_2015} is an emerging distributed learning framework. It allows multiple clients or participants (e.g., private mobile devices or IoT devices) to join the federated training collaboratively and improve model generalization without sharing their private data. Specifically, the clients can train the distributed model on their private datasets locally and submit the local model or gradient to the server, then the server will aggregate these updates using an aggregation algorithm (AGR). 

Recent studies have highlighted the vulnerability of federated learning (FL) to gradient inversion adversaries, who aim to reconstruct private local training data by exploiting the gradients shared during the FL process.
One common strategy employed by gradient inversion attacks is to optimize random noise in order to approximate the true gradient and subsequently reconstruct the original input data~\cite{fredrikson_model_2015,zhu_deep_2019,geiping_inverting_2020,yin_dreaming_2020,yin_see_2021,zhao_idlg_2020,wen_fishing_2022,fowl_robbing_2022,boenisch_when_2021,jeon_gradient_2021}. 
However, existing gradient inversion attacks (GIAs) against FL exhibit the following limitations:  
\begin{itemize}
    \item 
    Many existing GIAs necessitate control over the server to launch the attack~\cite{geiping_inverting_2020,yin_see_2021,wen_fishing_2022,fowl_robbing_2022,boenisch_when_2021}. However, in real-world scenarios, gaining control over the server is often challenging as it is typically heavily secured to prevent unauthorized access. 
    Moreover, if the AGR operates in the encrypted domain \cite{zhang2020batchcrypt}, all server-side existing GIAs would fail. 
    Therefore, client-side GIAs gain greater significance. However, the efficacy of such attacks~\cite{geiping_inverting_2020,zhu_deep_2019,yin_see_2021} remains limited to unrealistically small batch sizes. 
    \item Existing GIAs also rely on strong prior knowledge or specific information to achieve their objectives. For instance, GradInversion~\cite{yin_see_2021} requires access to batch normalization information, while generative-based methods~\cite {jeon_gradient_2021,usynin_beyond_2022,karras_analyzing_2020} utilize the distribution of the training data as prior knowledge for the adversary. These requirements for strong prior knowledge or specific information in GIAs restrict their applicability and effectiveness in real-world settings, where adversaries often have limited access to such privileged information. 
    \item Some GIAs necessitate altering the model structure or parameters, potentially leading to unintended consequences. For instance, some methods require the adversary to manipulate the training process in a way that traps clients into a predetermined status preferred by the adversary~\cite{boenisch_when_2021,wen_fishing_2022}. However, this alteration can negatively impact the main task performance or even disrupt the entire training framework~\cite{fowl_robbing_2022}.
\end{itemize}

{}


In this work, we propose the first~\cite{zhang2022survey} \textbf{C}lient-side poisoning \textbf{G}radient \textbf{I}nversion attack (\attack). In our \attack, the adversary first creates  a malicious model that amplifies the loss of a specific targeted class of interest on a local level. The adversary can then utilize this malicious model to poison the global model from the client side. When honest clients employ the poisoned global model, the gradients of samples belonging to the targeted class are magnified. As a result, the targeted gradient becomes the dominant factor in the aggregated update, allowing the adversary to effectively reconstruct the targeted private input belonging to other clients using the aggregated update. 
As such,  our \attack  is capable of successfully inverting gradients with larger batch sizes. Additionally, \attack is designed to function without any prior knowledge. Even if the adversary lacks information about the data distribution, they can still recover private data.  Furthermore, our \attack  aligns with the general training framework of FL and maintains the performance of the main task while carrying out the attack.



Another highlight of \attack is its ability to remain stealthy even when confronted with server-side defense measures, specifically Byzantine-robust AGRs which aim to identify and filter out malicious updates by analyzing the distribution of updates~\cite{blanchard_machine_2017,mhamdi_hidden_2018} and/or their performance on the validation dataset~\cite{cao_fltrust_2022,fang_local_2020}. We develop an effective method for optimizing malicious updates by disguising them as benign updates. This involves blending the benign update with a malicious replacement vector. The attacker can fine-tune the mixing hyper-parameters to accelerate the poisoning process or enhance the update's stealthiness against Byzantine-robust AGRs.

We also consider our \attack in a range of diverse FL settings where the adversary’s knowledge about data can  vary significantly. We consider three distinct levels of knowledge based on the data distribution information: i) \textit{Full Knowledge}: In this setting, the adversary possesses the knowledge of the data distribution across all classes. ii) \textit{Semi Knowledge}: The adversary only has access to the data distribution information specifically for the targeted class. iii) \textit{No Knowledge}: In this scenario, the adversary lacks any prior knowledge regarding the data distribution of any class. 

Our results demonstrate that \attack poses a high risk against FL. 
We conducted a comprehensive evaluation of \attack using three benchmark datasets and tested it against  four representative Byzantine-robust AGRs across various levels of knowledge. Our results show that \attack consistently and successfully extracts training input in all scenarios, surpassing the performance of existing GIAs. Notably, even in a scenario where the adversary possesses no prior knowledge, \attack  can successfully restore the training samples from batch sizes of 32 while the baselines only yield noise without any meaningful semantic features. 
Once given extra knowledge, a boost in performance is observed.

The contributions of our work are summarized as follows:
\begin{itemize}
    \item 
    We propose a new GIA, which is the first client-side gradient inversion attack capable of reconstructing another client's training sample from a large batch. It eliminates the strong assumption on the prior knowledge. Furthermore, \attack follows the general training framework without necessitating any modifications to the model structure or parameters and maintains normal performance on the main task at hand while executing the attack.  
    \item 
    Our \attack demonstrates the ability to evade detection by the Byzantine-robust AGRs. We reduce the attack into an optimization problem that  maximizes the adversarial impact on the victim training samples while ensuring that the malicious updates remain disguised. \attack can effectively generate malicious local updates 
    while remaining camouflaged within the normal training data. 
    \item 
    We evaluate our \attack under a comprehensive threat model including real-world attack scenarios considering different knowledge that adversary can obtain through FL. The results demonstrate the strength of the proposed \attack, showcasing its ability to succeed under different knowledge scenarios and reinforcing its effectiveness in practical settings.
    \item We release the source code and the artifact at \url{https://github.com/clientSideGIA/CGI}, 
    which creates a new tool for the GIA arsenal to facilitate future studies in this area.
\end{itemize}

\section{Background and Related Work}

\subsection{Federated Learning}
Federated learning (FL)~\cite{mcmahan_communication-efficient_2017,konecny_federated_2016,shokri_privacy-preserving_2015} is a distributed learning framework. Generally, the common FL algorithms are FedAvg~\cite{mcmahan_communication-efficient_2017} and FedSGD~\cite{mcmahan_communication-efficient_2017}. The standard FL structure contains a server model $\Phi_G$ and $N$ clients. Each client $i$ has a private dataset $D_i$ drawn from a data distribution $\mathcal{D}$.

For FedSGD, in every round $t$, each client $i$ first receives the global model $\Phi_G^{t-1}$ broadcasted by the server, then commits the gradient $\delta_i^t = \nabla_{\Phi_G^{t-1}}\mathcal{L}(D_{i})$ to the server. The server will aggregate selected client gradients and use the averaged gradient $\overline{\delta}^t = \frac{1}{N}\sum_{i=1}^N\delta_i^t$ to update the global model and obtain $\Phi_G^t$. Then, the server distributes $\Phi_G^t$ to each client and begins the next training round.

For FedAvg, it is similar to FedSGD, but each client $i$ can initialize a local model $\dot{\Phi}_i^{t}$ with $\Phi_G^{t-1}$ and run more local epochs with dataset $D_i$ to optimize the local model $\dot{\Phi}_i^{t}$. The server will collect all trained local models and aggregate them with weight averaging, then the server gets a new global model $\Phi_G^t = FedAvg(\{\dot{\Phi}_i^t|i\in[N]\})$. Note that the model update in FedAvg equals the gradient update in FedSGD when FedAvg only runs one local epoch. Then the server will distribute the current global model to every client and begin the next round. The key feature of FL is that no private data will be disclosed to other clients or the server. 

\subsection{Byzantine-robust Aggregation Rules}
Due to the distributed nature of FL, 
multiple Byzantine-robust AGRs~\cite{blanchard_machine_2017,mhamdi_hidden_2018,cao_fltrust_2022,fang_local_2020,xie_generalized_2018,yin_byzantine-robust_2018} are proposed to defend against malicious updates from clients. \textit{Krum}, \textit{Multi-krum}~\cite{blanchard_machine_2017} and \textit{Bulyan}~\cite{mhamdi_hidden_2018} 
aim to identify and remove malicious gradients by calculating the distance between different updates. The idea is to cluster similar updates together and treat the remaining gradients as potentially malicious. 
\textit{Trimmed-mean} and \textit{Median}~\cite{xie_generalized_2018,yin_byzantine-robust_2018}
use either the median or trimmed mean (excluding extreme values) to aggregate gradients, aiming to mitigate the influence of extreme gradients. 
\textit{AFA}~\cite{cao_fltrust_2022}
introduces a trust bootstrapping mechanism to identify and eliminate potentially harmful local updates that deviate significantly from a trusted clean dataset used as a reference. 
\textit{Fang defense}~\cite{fang_local_2020} 
uses a validation set on the server side to filter gradients. It evaluates the performance of gradients on the validation set and discards those that do not align with the overall model's behavior.


{}

\subsection{Local Model Poisoning Attack against FL}
Model poisoning attacks against FL systems can have two objectives: breaking model performance to cause Denial-of-Service (DoS) or introducing a backdoor into the model. 

When aiming to disrupt the performance of a model, recent studies have focused on developing  strategies to create local poisoned models or model updates that can successfully penetrate Byzantine-robust AGRs.
An example of such an attack is the LIE attack~\cite{baruch_little_2019}, which requires minimal knowledge of Byzantine-robust AGRs and introduces small perturbations to benign gradients to evade detection. 
Fang~\cite{fang_local_2020} introduced a general optimization problem where adversaries optimize local poisoned gradients to make the aggregated global model deviate in the opposite direction from the previous round. 
Recently, Shejwalkar et al.~\cite{shejwalkar_manipulating_2021} proposed a comprehensive framework for model poisoning under robust AGRs, considering different levels of knowledge and developing AGR-tailored attacks and AGR-agnostic attacks.


In the case of backdooring FL systems, Bagdasaryan et al.~\cite{bagdasaryan_how_2020} proposed a method where an adversary trains a backdoor model locally and subsequently replaces the global model by employing adversarial model replacement methods. 
Wang et al.~\cite{wang_attack_2020} demonstrated the feasibility of utilizing edge-case samples to introduce a backdoor into the global model. 
Bhagoji et al.~\cite{bhagoji_analyzing_2019} provided an effective optimization process for executing a backdoor attack under Byzantine-robust AGRs in FL. 
This approach incorporates a stealthy element that optimizes malicious gradients to align with the average of benign gradients. 

Different from traditional poisoning attack which aim to corrupting the model's integrity, \attack's objective is to reconstruct the original private training samples through poisoning. 



\subsection{Inversion Attack} \label{section-literature-inversionattck}

Inversion attacks typically consist of two types: model inversion attack (MIA) and gradient inversion attack (GIA). In these attacks, the adversary aims to reconstruct original training data by utilizing either gradient information or a publicly available model.


Typical MIA~\cite{fredrikson_model_2015} attempts to fit real confidence score distribution to extract training data from DNN based on black-box access. More advanced model inversion attack can only use label~\cite{kahla_label-only_2022} or exploit GAN~\cite{struppek_plug_2022} to launch the attack.

GIA usually poses a greater threat compared to MIA due to its ability to reconstruct user's private input with higher precision. The adversary's goal in a GIA attack is to find an input $(x, y)$ such that it can produce the same gradient as a certain local gradient update (or equivalently model update), which can be observed by the server during FL training. It is generally achieved by optimizing the $l_2$ distance~\cite{zhu_deep_2019} or cosine similarity~\cite{geiping_inverting_2020} between the gradient associated with $(x, y)$ and server's observation. Inferring the true value of $y$ (i.e., label inference) is essential in the effectiveness of GIAs, including reducing the search space and stabilizing the optimization processing, which finally improves inversion performance. In particular, iDLG~\cite{zhao_idlg_2020} and GradInversion~\cite{yin_see_2021} developed successful zero-shot label inference methods based on the feature of cross-entropy loss and the final layer gradient. By assuming knowledge of batch normalization layers as a prior, GradInversion~\cite{yin_see_2021} makes an amazing effect on image gradient inversion. Some researchers also demonstrated other methods to invert gradients besides optimization methods. One of the successful methods based on mathematical analysis to infer input~\cite{zhu_r-gap_2022}. In such attacks, the adversary generally plays as server side and sends malicious parameters to clients~\cite{boenisch_when_2021,wen_fishing_2022} or change framework structure~\cite{fowl_robbing_2022} to capture some features. Another successful method is based on generative models. These methods ~\cite{jeon_gradient_2021,usynin_beyond_2022,karras_analyzing_2020} can produce realistic data, but the data is different from the inputs. 

\section{Threat Model}
\subsection{Adversary’s Capabilities} 
We consider a typical poisoning client-side adversary's capability in a common federated learning system based on \textit{FedSGD}~\cite{mcmahan_communication-efficient_2017} 
following the prior research~\cite{bagdasaryan_how_2020, bhagoji_analyzing_2019, cao_fltrust_2022, shejwalkar_manipulating_2021, shen2022better}. 
The adversary can download the global model of each round and take the full control of $m$ out of $N$ clients, including their local updates and local training data. The adversary can also decide whether to inject malicious updates for the current round or not.
We assume the adversary does not compromise the central aggregator.  
The server can equip the defense 
methods, i.e., \textit{Byzantine-robust AGRs
}.





\subsection{Adversary’s Goal}
The adversary aims at restoring the training samples that he/she interests (i.e., he/she attacks samples of any one class at a time). We call the sample that is under the current round of attack as the \emph{targeted sample}, and the class that the targeted sample belongs to as the \emph{targeted class}. For the other samples that are not under attack, we refer to them as the \emph{untargeted sample} and their corresponding classes as the \emph{untargeted class}.  


\subsection{Adversary’s Knowledge}\label{sec:knowledge}
We place \attack under diverse FL settings with varying levels of adversary knowledge about the data, so as to  conduct a comprehensive evaluation of \attack. 
We consider the adversary's knowledge from two aspects:  knowledge of training data distribution and knowledge of the AGR algorithm of the server.

For the knowledge of data distribution (denote as $\mathcal{K}$), we consider three levels: 
\begin{itemize}
    \item \emph{Full Knowledge ($\mathcal{K}_{full}$).} The adversary has the full knowledge of the data distribution of all classes. As such, he/she can generate synthetic training records of each class from their marginal distributions. Such an assumption applies to the FL with IID data in which the adversary is able to obtain the statistical information of all classes from its own training dataset. 
    \item \emph{Semi Knowledge ($\mathcal{K}_{semi}$).} The adversary has the knowledge of the data distribution of the targeted class, and the knowledge of the data distribution of some other (not full) untargeted classes. This assumption applies to the FL with Non-IID data in which the adversary only has the training samples from some certain classes including the targeted class. 
    \item \emph{No Knowledge ($\mathcal{K}_{no}$).} The adversary knows nothing about the underlying distribution of any classes. 
    This assumption also applies to the Non-IID FL, while in this case, the training samples held by the adversary are insufficient for them to derive the underlying distribution. In addition, the adversary does not hold any training samples from the targeted class.
    
\end{itemize}

For the knowledge of AGR,  we assume the adversary is AGR-agnostic, which is in favor of the defense mechanism as it makes the attack harder. 


\begin{figure}
    \centering
    \includegraphics[width=0.5\textwidth]{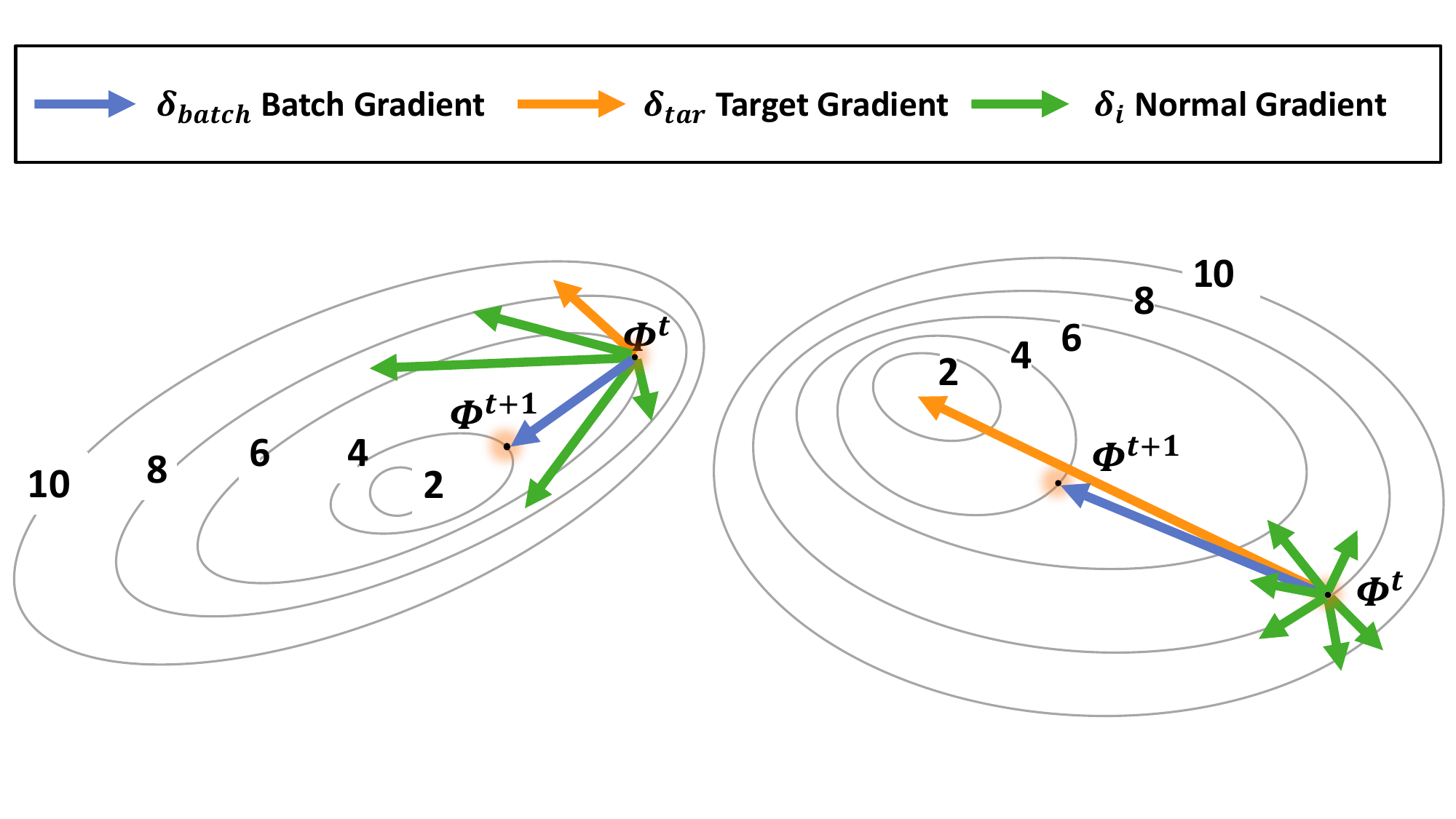}
    \caption{\textbf{Left}: gradient aggregation on normal local updates; \textbf{Right}: gradient aggregation with poisoned model.}
    \label{fig:intuition}
\end{figure}

\section{\attack}
\label{Sec:CTGIAttack}

\subsection{Intuition}
\label{subsection:Intuition}
The particular challenges for a client-side adversary are two-fold: (1) How to extract the targeted gradient from a batch without the strong assumption of prior knowledge,  
such as batch normalization (BN) information and data distribution? 
(2) How to retain the attack's effectiveness in the presence of AGRs? 
In the following, we present the insight we use in \attack to tackle these challenges. 

\paragraph{Obtain the targeted gradient from the batch} 
We design a poisoning client-side adversary that can actively influence the global model in order to extract the gradient of the targeted sample from batches. 

In \attack, we let the adversary push updates towards the malicious direction of the targeted sample's gradients. When the data owner runs its local gradient descent, this malicious updates that is injected into the global model will cause an abrupt reduction of the gradient on the targeted sample while the gradients of other untargeted samples are kept unchanged in the batch. The adversary can thus exploit such differences to obtain the targeted sample's gradients. 
Particularly, when the untargeted samples' gradients are small, the batch gradients will be dominated by the targeted sample's gradients. This further eases the attack as it allows the adversary to conduct the gradient inversion on the batch gradients directly to restore the targeted sample. 

To illustrate this, let $\Phi_G$ denote the global parameters of a DNN model and  and $\mathcal{L}$ denote the loss function. Let $(X,Y)$ be a batch of data samples with $X = \{x_i\}_{i=1}^{|B|}$ and $Y = \{y_i\}_{i=1}^{|B|}$. The batch gradient $\delta_{batch}$ is
$$\delta_{batch} = \frac{1}{|B|}\sum_{i=1}^{|B|}\nabla_{\Phi_G}\mathcal{L}((x_i,y_i),\Phi_G).$$

Let $x_{tar}$ denote the targeted sample that belongs to the targeted class ${tar}$. If the gradient $\delta_{tar}$ of the loss with respect to parameters on $x_{tar}$ is far larger than the gradient on other samples in the same batch, that is: 
$$|\nabla_{\Phi_G}\mathcal{L}((x_{tar},y_{tar}),\Phi_G)| \gg \sum_{y_j\neq y_{tar}}|\nabla_{\Phi_G}\mathcal{L}((x_j,y_j),\Phi_G)|.$$  
Then the gradient $\delta$ of loss $\mathcal{L}$ w.r.t. model parameters on batch data is:
$$\delta_{batch} \approx \frac{1}{|B|} \nabla_{\Phi_G}\mathcal{L}((x_{tar},y_{tar}),\Phi_G).
$$
Therefore, the objective of the adversary is to poison  $\Phi_G$ such that it maximizes the difference between the targeted gradient and the others, that is: 
\vspace{-0.2cm}
\begin{equation}
\label{equ:opt_mal_model}    
\small{
\arg\max\limits_{\Phi_G}\left(|\nabla_{\Phi_G}\mathcal{L}((x_{tar},y_{tar}),\Phi_G)| - \sum_{y_j\neq y_{tar}}|\nabla_{\Phi_G}\mathcal{L}((x_j,y_j),\Phi_G)|\right). 
}
\end{equation}

\normalsize
Figure~\ref{fig:intuition} demonstrates differences between gradients on a normal model and gradients on a poisoned model, where $t$ is the training round. {\iffalse$\delta_{tar}$ is the targeted sample's gradient and $\delta_{i}$ is any non-targeted sample's gradient. $\delta_{batch}$ represents the batch gradient which is produced by averagely aggregating each sample's gradient in the batch.\fi}  
In Figure~\ref{fig:intuition} Left, $\delta_{tar}$ is similar to any $\delta_{i}$, and has no obvious relationship with $\delta_{batch}$. However, in Figure~\ref{fig:intuition} Right,  $\delta_{tar}$ can dominate $\delta_{batch}$ on a targeted poisoning model because the $|\delta_{tar}|$ is far larger than $\sum_{j=1, y_j \neq y_{tar}}^{|B|}|\delta_j|$ so that the adversary can directly use $\delta_{batch}$ as $\delta_{tar}$. 

\begin{figure*}[t]
\centering
\includegraphics[width=0.8\linewidth]{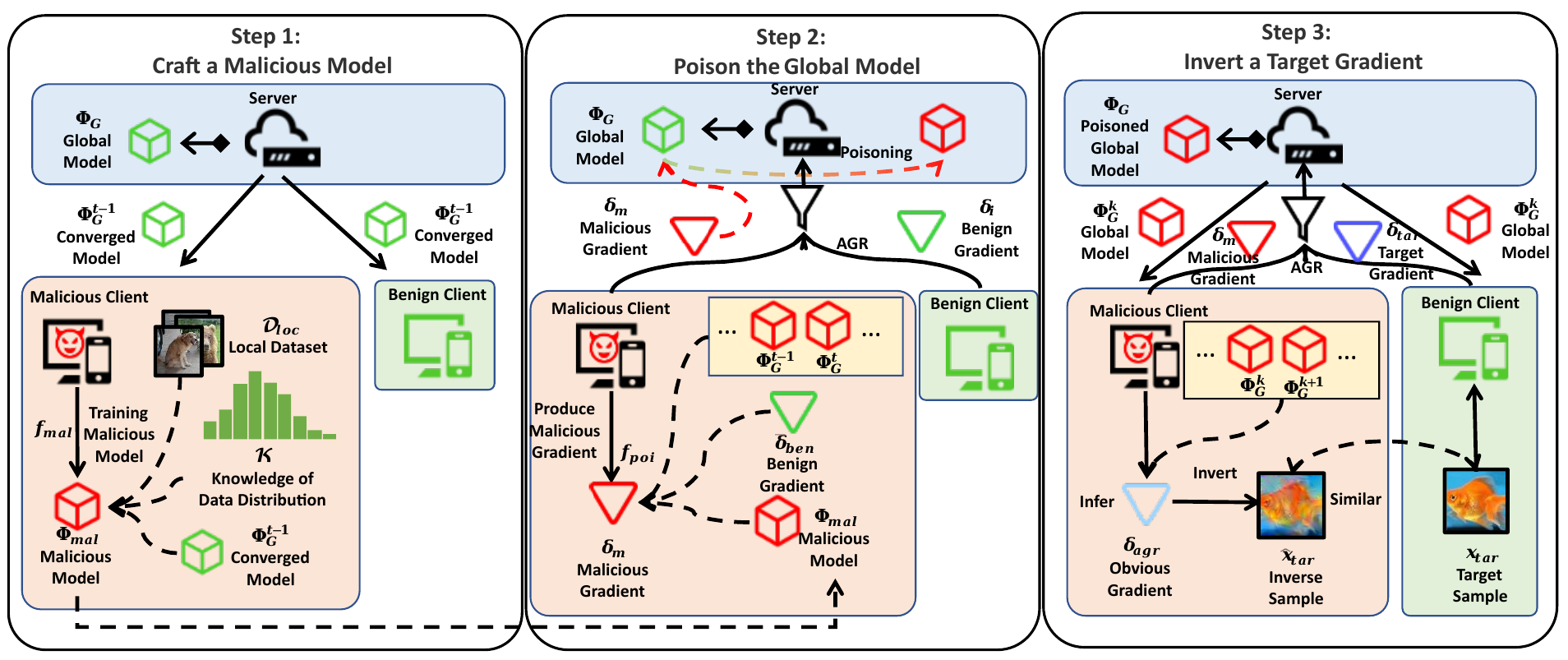}
\caption{Overview of \attack.}
\label{fig:overview}
\end{figure*}


\paragraph{Evade Byzantine-robust AGRs} 
To evade Byzantine-robust AGRs, we need the malicious updates crafted by the adversary to be close to the updates of the honest clients. 
To this end, we reduce our \attack into an optimization problem 
that minimizes the distance between the to-be-crafted malicious updates and the honest clients' updates by covering the maliciousness with an add-on benign gradient while retaining the adversarial impact with respect to the targeted sample. 
Once the attack successfully evades detection, the malicious impact on the targeted samples will take effect and lead to successful client-side gradient inversion.

\subsection{Overview of \attack}

Figure~\ref{fig:overview} presents the overview of \attack, which consists of three steps. 
\paragraph{Step 1. Craft a malicious model locally}
In this step, the adversary aims to craft a local malicious model that maximizes the loss of the targeted sample, that is:
\begin{equation}
    \Phi_{mal} = f_{mal}(\Phi_G^{t-1},\mathcal{K},D_{loc}),
\end{equation}
where $\Phi_G^{t-1}$ is the global model at round $t-1$ (before poisoning), $\mathcal{K}$ represents the knowledge of training data distribution (aforementioned in Section~\ref{sec:knowledge}) and $D_{loc}$ denotes a local training dataset held by the adversary. 
The function $f_{mal}$ outputs $\Phi_{mal}$ which performs normally on untargeted samples but fails on the targeted samples.  
We propose different $f_{mal}$ for various  $\mathcal{K}$ accordingly for \emph{full knowledge}, \emph{semi knowledge} and \emph{no knowledge} of data distribution, respectively, detailed later in Section~\ref{sec:mal}. 

\paragraph{Step 2. Poison the global model}
In this step, the adversary aims to submit a malicious update $\delta_m$  that contains the maliciousness of $\Phi_{mal}$ while breaking the AGRs. This can be characterized by 
\begin{equation}
    \delta_m = f_{poi}(\Phi_{G}^{t-1}, \Phi_{G}^t,\Phi_{mal},\overline{\delta}_{ben}),
\end{equation}
where $\overline{\delta}_{ben}$ represents the average of benign gradients from the malicious clients, and  $\Phi_{G}^{t-1}$ and $\Phi_{G}^t$ are used to approximate the aggregated gradients of the server in the absence of adversary. 
$f_{poi}$ denotes an optimization process that minimizes the difference of $\delta_m$ and other clients' updates while retaining the effect of $\Phi_{mal}$. 
The adversary can produce and submit the malicious update in multiple training rounds until the global model has been successfully poisoned. 
Detailed construction of $\delta_m${\iffalse$\overline{\delta}_{ben}$\fi} and the optimization process are given in Section~\ref{sec:poi}.

\paragraph{Step 3. Invert the targeted gradient}
The final step is to invert gradients of the targeted sample from its owner.
The adversary uses a gradient inversion function to restore the targeted sample as: 
\begin{equation}
    \hat{x}_{tar} = f_{inv}(\Phi_G^{k},\Phi_G^{k+1}, \cdots), 
\end{equation}
where $(\Phi_G^{k},\Phi_G^{k+1}, \cdots)$ denote the poisoned global models from round $k$ ($k \geq t$) onwards. If a certain training round contains $x_{tar}$, its gradients will dominate the aggregated updates (i.e., the difference between the global model of the current and the previous round). This then enables the adversary to restore the targeted sample from the aggregated updates. 
Detailed explanation of 
$f_{inv}$ are given in Section~\ref{sec:inv}. We also provide symbol tables in~\ref{appendix:symbol_tables}.

\subsection{Craft the Malicious Model}\label{sec:mal}
In this section, we discuss the methods of crafting the malicious model based on the adversary's different levels of knowledge.

\begin{algorithm}[!htbp]
\caption{Crafting the Mal. Model with Full Knowledge.}\label{alg:MR}
\begin{algorithmic}[1]
\Require 
\Statex $\mathcal{K}_{full}$: the full knowledge of the data distribution of all classes;
$\Phi^0$: the initialized model;
$r_{end}$: the number of local training rounds; 
$\alpha_1$: the normal training learning rate; 
$\alpha_2$: the trap training learning rate;
\Ensure 
\Statex $\Phi_{mal}$: the targeted poisoning model
\State Construct $D_{aux}\sim\mathcal{K}_{full}$
\State Construct $D_{main}=\{(x_i,y_i)\}$ from $D_{aux}$
\State Construct $D_{trap}=\{(x_i,y_{tar})\}$ from $D_{aux}$
\State Initialize {$\Phi^{r_0} \leftarrow \Phi^{0}$}
\For{$r \leftarrow 0$ to $r_{end}$}
    \Statex \qquad  \textcolor{blue}{\(\triangleright\) Minimize main task loss.}
    \State $\tilde{\Phi}^{r+1} = \Phi^r - \alpha_{1}\nabla_{\Phi^r}\mathcal{L}(\Phi^r,D_{main})$ 
    \Statex \qquad \textcolor{blue}{\(\triangleright\) Maximize targeted class loss.}
    \State $\Phi^{r+1} = \tilde{\Phi}^{r+1} + \alpha_{2}\nabla_{\tilde{\Phi}^{r+1}}{\mathcal{L}(\tilde{\Phi}^{r+1}},D_{trap})$ \
\EndFor
\State $\Phi_{mal} \leftarrow \Phi^{r_{end}}$ \\
\Return $\Phi_{mal}$

\end{algorithmic}
\end{algorithm}

\subsubsection{Full Knowledge ($\mathcal{K}_{full}$)}
In this setting, the adversary does not require the global model from the server and all operations can run locally. He/She can directly 
synthesize an auxiliary dataset $D_{aux}$ that has the same distribution as $\mathcal{K}_{full}$. Then $D_{aux}$ can represent the entire data distribution, thus $D_{loc}$ can be ignored.

The adversary first removes the samples of the targeted class from $D_{aux}$. We denote this dataset without the targeted class as $D_{main}$. Training on $D_{main}$ can ensure the local model to achieve the normal performance on the main task while learning nothing about the targeted class. 

To further enlarge the performance gap between the targeted class and the main task, the adversary additionally prepares a dataset called  $D_{trap}$ which consists of the samples from $D_{main}$ but falsely labeled by the targeted class $y_{tar}$. We conduct gradient ascent when training on $D_{trap}$. As such, the loss with respect to the targeted class is significantly increased as the local model is updated towards the opposite direction of predicting an arbitrary sample to the targeted class (that is, the local model will never classify any samples to the targeted class). In addition, such training on $D_{trap}$ in the gradient ascent manner will not affect the performance on the main task as well.

 To craft the malicious model, the adversary iteratively trains on  $D_{main}$ and $D_{trap}$ for $r_{end}$ rounds that the adversary can set, as shown in Algorithm~\ref{alg:MR}. 



\setlength{\parskip}{0pt}
\subsubsection{Semi Knowledge ($\mathcal{K}_{semi})$}\label{sec:semi}
In this setting, the adversary first pulls the nearly converged global model $\Phi_G^{t-1}$ from the server (the adversary can wait for sufficient training rounds or estimate global model convergence based on the local validation dataset collected), then he/she is able to generate a $D_{tar}$ that matches the distribution of the targeted class only based on $\mathcal{K}_{semi}$. To craft the malicious model, the adversary first calculates the gradients with respect to the targeted class based on $D_{tar}$. Then he/she pushes the local updates towards the ascending direction of the targeted gradients (also known as the targeted model poisoning~\cite{munoz-gonzalez_towards_2017}). 

However, such loss on the targeted class will damage model generalization on the main task to some extent. Therefore, we let the adversary construct a $D_{ben}$ from its own local datasets $D_{loc}$. The $D_{ben}$ only contains the adversary's local samples for the main task, i.e., the samples of the targeted class are removed. He/she then uses it to repair the lost cross-class features. The algorithm has been shown in Algorithm~\ref{alg:MP}. 

\begin{algorithm}[!htbp]
\caption{Crafting the Mal. Model with Semi Knowledge.}\label{alg:MP}
\begin{algorithmic}[1]
\Require 
\Statex $\mathcal{K}_{semi}$: the semi-knowledge of the data distribution of the targeted class;
$D_{loc}$: the local dataset;
$\Phi_{G}^{t-1}$: the converged global model;
$r_{end}$: the number of local training rounds; 
$\alpha_1$: the poisoning training learning rate; 
$\alpha_2$: the repair training learning rate;
\Ensure 
\Statex $\Phi_{mal}$: the targeted poisoning model
\State Construct $D_{tar}=\{(x_{tar},y_{tar})\} \sim \mathcal{K}_{semi}$
\State Construct $D_{ben}=\{(x_i,y_i)\}$ from $D_{loc}$
\State Initialize {$\Phi^{r_0} \leftarrow \Phi_{G}$}
\For{$r \leftarrow 0$ to $r_{end}$}
    \Statex \qquad  \textcolor{blue}{\(\triangleright\) Poisoning targeted class.}
    \State $\tilde{\Phi}^{r+1} = \Phi^r + \alpha_{1}\nabla_{\Phi^r}\mathcal{L}(\Phi^r,D_{tar})$ 
    \Statex \qquad \textcolor{blue}{\(\triangleright\) Repair main task performance.}
    \State $\Phi^{r+1} = \tilde{\Phi}^{r+1} - \alpha_{2}\nabla_{\tilde{\Phi}^{r+1}}{\mathcal{L}(\tilde{\Phi}^{r+1}},D_{ben})$ \
\EndFor
\State $\Phi_{mal} \leftarrow \Phi^{r_{end}}$ \\
\Return $\Phi_{mal}$

\end{algorithmic}
\end{algorithm}
\setlength{\textfloatsep}{0pt}


\setlength{\parskip}{0pt}
\subsubsection{No Knowledge ($\mathcal{K}_{no}$)} 
In this setting, the adversary does not have any samples of the targeted class. 
In order to craft the desired malicious model against the targeted class, the adversary must first obtain some knowledge about it. 
To this end, we let the adversary conduct model inversion~\cite{fredrikson_model_2015,ye_model_2022,yin_dreaming_2020} against the coveraged global model $\Phi_{G}^{t-1}$, aiming to reconstruct some samples from the targeted class by solving
\begin{equation}
\begin{aligned}
\hat{X}_{tar} = \arg\min\limits_{\hat{X}}& \ \lambda_0 \mathcal{L}(\Phi_G^{t-1},(\hat{X},Y_{tar})) 
\\ 
&+ \lambda_1 \mathcal{R}_{TV}(\hat{X}) 
+\lambda_2 \mathcal{R}_{l_2}(\hat{X})
+\lambda_3 \mathcal{R}_{f}(\hat{X}),
\end{aligned}
\end{equation}
where $\hat{X}_{tar}$ represents reconstructed samples of targeted class, $Y_{tar}$ is the targeted class label array, and $\lambda_i$ ($i\in \{0,1,2,3\}$) is the coefficient. Here, $\mathcal{R}_{TV}$ and $\mathcal{R}_{l_2}$ are loss functions that regularize the total variation and $l_2$ norm of $\hat{X}$ and $\mathcal{R}_f$ is to ensure feature similarities between the mean and  variance of $\hat{X}$ and the global model's batch normalization layer, which is formalized as:
\begin{equation}
\begin{aligned}
\mathcal{R}_{f}(\hat{X}) & = \sum\limits_l||\mu_l(\hat{X})-BN_l(mean)||_2 \\
&+ \sum\limits_l||\sigma^2_l(\hat{X})-BN_l(variance)||_2.
\end{aligned}
\end{equation}
Here we point out that the required batch normalization layer information is different from those required by GradInverion~\cite{yin_see_2021}. GradInverion~\cite{yin_see_2021} requires batch-wise mean and variance of local models but our method only needs the batch normalization layer information from the global model and such information is insensitive to specific batch input.


We note that due to the limited knowledge owned by the client-side adversary, the reconstructed samples $\hat{X}_{tar}$ in this step can be far away from the real training samples: they can be vague, distorted and contain many noises. However,  the information about the common features  contained in $\hat{X}_{tar}$ can be utilized by the adversary to perturb the local model towards the malicious direction against the targeted class. 

Next, following the method in Section~\ref{sec:semi} (i.e., Semi knowledge), the adversary constructs the malicious model by ascending the targeted gradients based on $\hat{D}_{tar}$ constructed by $\hat{X}_{tar}$ and repairs the performance of the main task using his/her local datasets $D_{loc}$ (no targeted samples because of $\mathcal{K}_{no}$). The training algorithm has been shown in Algorithm~\ref{alg:MIP}.  

\begin{algorithm}[t]
\caption{Crafting the Mal. Model with No Knowledge.}\label{alg:MIP}
\begin{algorithmic}[1]
\Require 
\Statex $D_{loc}$: the local dataset based on no knowledge; 
$\Phi_{G}^{t-1}$: the converged global model;
$p_{end}$: the number of model inversion rounds; 
$r_{end}$: the number of local training rounds; 
$\lambda_i~(i \in \{0,1,2,3\})$: the coefficients in model inversion; $\alpha_1$: the poisoning training learning rate; 
$\alpha_2$: the repair training learning rate;
\Ensure 
\Statex $\Phi_{mal}$: the targeted poisoning model
\State Initialize {$\hat{X}_{tar}$}
\State Define $\mathcal{L}_{MI}(\hat{X}) = \lambda_0 \mathcal{L}(\Phi_G,(\hat{X},Y_{tar})) 
    + \lambda_1 \mathcal{R}_{TV}(\hat{X}) 
    + \lambda_2 \mathcal{R}_{l_2}(\hat{X})
    + \lambda_3 \mathcal{R}_{f}(\hat{X})$
\Statex \textcolor{blue}{\(\triangleright\) Inverting global model to obtain target samples.}
\For{$p \leftarrow 0$ to $p_{end}$}
    \State $\hat{X}_{tar}^{p+1} = \hat{X}_{tar}^{p} - \nabla_{\hat{X}_{tar}^{p}}\mathcal{L}_{MI}(\hat{X}_{tar}^{p})$
\EndFor
\State $\hat{X}_{tar} \leftarrow \hat{X}_{tar}^{p_{end}}$
\State Construct $\hat{D}_{tar} = \{(\hat{x}_{tar},y_{tar})\}$
\State Initialize {$\Phi^{r_0} \leftarrow \Phi_{G}$}
\For{$r \leftarrow 0$ to $r_{end}$}
    \Statex \qquad  \textcolor{blue}{\(\triangleright\) Poisoning target class.}
    \State $\tilde{\Phi}^{r+1} = \Phi^r + \alpha_{1}\nabla_{\Phi^r}\mathcal{L}(\Phi^r,\hat{D}_{tar})$ 
    \Statex \qquad \textcolor{blue}{\(\triangleright\) Repairing main task performance.}
    \State $\Phi^{r+1} = \tilde{\Phi}^{r+1} - \alpha_{2}\nabla_{\tilde{\Phi}^{r+1}}{\mathcal{L}(\tilde{\Phi}^{r+1}},D_{loc})$ \
\EndFor
\State $\Phi_{mal} \leftarrow \Phi^{r_{end}}$\\
\Return $\Phi_{mal}$
\end{algorithmic}
\end{algorithm}
\setlength{\textfloatsep}{0pt}

\subsection{Poison the Global Model}\label{sec:poi}
In this section, we show the process of replacing the global model with the crafted malicious model while defeating the AGRs. 

We define the replacement problem below. 
Let 
$$\delta_{poi} = \Phi_G^t - \Phi_{mal}$$ 
be the updates that the adversary aims to inject into the global model. Clearly, if there is no AGR equipped, the adversary can directly commit $\delta_{poi}$ as his malicious update $\delta_m$ to boost attack efficiency.

{} 
{}
{}

In order to evade the AGRs, the adversary needs to minimize the distance between $\delta_{m}$ and the current round aggregated gradient $\delta_{agr}$. Thus, the adversary's objective becomes
\begin{equation}
\begin{aligned}
 \arg\min\limits_{\delta_m} f_{dis}(\delta_{m}, \delta_{agr}),~\label{equ:agr}
\end{aligned}
\end{equation}
where $f_{dis}$ is a distance metric function.
%
However, the above formula is hard to solve due to the variety of AGRs and the adversary cannot obtain the current round aggregated gradient $\delta_{agr}$. Inspired by the literature of optimizing a traditional poisoning attack~\cite{fang_local_2020,shejwalkar_manipulating_2021,bhagoji_analyzing_2019}, we express it in a different form that is better suited for optimization and universal for all AGRs. 

{}

The adversary first calculates the aggregated updates of the previous round as 
$$\delta_{agr} = \frac{1}{\alpha}(\Phi_G^{t-1} - \Phi_G^{t}).$$  
Then, he/she constructs $\delta_m$ as 
$$\delta_m = \gamma\frac{\delta_{poi}}{||\delta_{poi}||} + \beta \overline{\delta}_{ben},$$
where $\overline{\delta}_{ben}$ represents the average of benign gradients from malicious clients (i.e., calculated by malicious clients on their local datasets). It can be calculated from the local dataset $D_{loc}$ of every malicious client. The parameters $\gamma$, $\beta$ are the scaling coefficients for the unitary malicious gradient and reference gradient, which is initialized as $\gamma_0$ and $\beta_0$ (and they are tunable).
Therefore, Equation~(\ref{equ:agr}) can be transformed as: 
\begin{equation}
\begin{aligned}
\arg\min\limits_{\gamma,~\beta}||\delta_{m} - \delta_{agr}||_2, 
\end{aligned}
\end{equation}
by which the adversary minimizes the distance between malicious update $\delta_m$ and aggregated gradients $\delta_{agr}$ to break AGRs. Then the adversary still uses SGD to optimize $\gamma, \beta$. The optimization of the malicious update has been described in Algorithm~\ref{alg:OptMalUpdate}.

\begin{algorithm}[H]
\caption{Optimizing Malicious Update}\label{alg:OptMalUpdate}
\begin{algorithmic}[1]
\Require 
\Statex $\Phi_{G}^t$: the global model at $t$ round;
$\Phi_{G}^{t-1}$: the global model at $t-1$ round;
$\Phi_{mal}$: the malicious model;
$\overline{\delta}_{ben}$: the benign gradient;
$\gamma_0$: the initialized coefficient of malicious gradient; $\beta_0$: the initialized coefficient of benign gradient; $r_{end}$: the number of optimization rounds; 
$\alpha$: the learning rate of the server; 
$\alpha_1$: the learning rate for $\gamma$; 
$\alpha_2$: the learning rate for $\beta$;
\Ensure 
\Statex $\delta_{m}$: the malicious update
\Statex \textcolor{blue}{\(\triangleright\) Calculating the aggregation gradient of the last round.}
\State $\delta_{agr} = \frac{1}{\alpha}(\Phi_G^{t-1} - \Phi_G^t)$
\Statex \textcolor{blue}{\(\triangleright\) Optimizing the malicious update $\delta_m$.}
\State Initialize {$\gamma^{r_0} \leftarrow \gamma_0$}, {$\beta^{r_0} \leftarrow \beta_0$}
\State Initialize {$\delta^{r_0}_m \leftarrow \gamma^{r_0}\frac{\delta_{poi}}{||\delta_{poi}||} + \beta^{r_0} \overline{\delta}_{ben}$}
\State Define {$\mathcal{L}_{dis}(\delta_m) = ||\delta_m - \delta_{agr}||_2$}
\For{$r \leftarrow 0$ to $r_{end}$}
    \State $\gamma^{r+1} = \gamma^{r} - \alpha_1 \nabla_{\gamma^{r}}\mathcal{L}_{dis}(\delta^{r}_m)$
    \State $\beta^{r+1} = \beta^{r} - \alpha_2 \nabla_{\beta^{r}}\mathcal{L}_{dis}(\delta^{r}_m)$
    \State $\delta^{r+1}_m = \gamma^{r+1}\frac{\delta_{poi}}{||\delta_{poi}||} + \beta^{r+1} \overline{\delta}_{ben}$
\EndFor
\State $\delta_{m} \leftarrow \delta^{r_{end}}_{m}$\\
\Return $\delta_{m}$
\end{algorithmic}
\end{algorithm}


\subsection{Invert the Targeted Gradient}\label{sec:inv}
In this section, we describe the method of restoring the targeted sample from the poisoned global models. 

The adversary first pulls the poisoned global models $(\Phi_G^{k},\Phi_G^{k+1}, \cdots)$ from round $k$ ($k \geq t$) onwards. He/she then calculates the aggregated updates of each round as:  
$$\delta_{agr}^{k} = \frac{1}{\alpha}(\Phi_G^{k+1} - \Phi_G^k).$$ 

Next, the adversary conducts gradient inversion~\cite{geiping_inverting_2020, zhu_deep_2019} against each of $\delta_{agr}^{k}$ as follows. 
First, the adversary initializes the targeted sample $\hat{x}_{tar}$ with random noise and constructs dummy gradients with  
\begin{equation}
    \hat{\delta}_{tar} = \nabla_{\Phi_G^{k}}\mathcal{L}(\hat{x}_{tar},y_{tar}).
\end{equation}
Then the adversary aims to restore $\hat{x}_{tar}$ by solving: 
\vspace{-0.1cm}
\begin{equation}
\label{equ:inversion}
\begin{aligned}
\arg\min\limits_{\hat{x}_{tar}} &\ \lambda_0 (1-\frac{<\hat{\delta}_{tar},\delta_{agr}^{k}>}{||\hat{\delta}_{tar}||||\delta_{agr}^{k}||})
\\
&+ \lambda_1 R_{TV}(\hat{x}_{tar})
+ \lambda_2 R_{l_2}(\hat{x}_{tar})
+ \lambda_3 R_{clip}(\hat{x}_{tar}), 
\end{aligned}
\end{equation}
by which the adversary finds an optimal $\hat{x}_{tar}$ that maximizes the cosine similarity of $\hat{\delta}_{tar}$ and $\delta_{agr}^{k}$. 
The parameter $\lambda_i$ ($i \in \{0, 1, 2, 3\}$) is the coefficients, and $R_{TV}$, $R_{l_2}$ and $R_{clip}$ are the regularization terms of the total variation, the $l_2$ norm and the clip item, respectively~\cite{geiping_inverting_2020}. 

\section{Experimental Settings}


\subsection{Datasets and Model Architecture}
We use Cifar100~\cite{krizhevsky2009learning}, TinyImageNet~\cite{tiny-imagenet}, and CalTech256~\cite{griffin2007caltech} as benchmark datasets and choose Resnet18~\cite{he2016deep} as the model structure.

{}

\subsection{Implementation Settings}

\paragraph{FL settings} We set the number of clients $N$ varying from 10 to 50 in our experiments. The proportion of malicious clients is set to around 20\% following the common setting in prior research~\cite{shejwalkar_manipulating_2021,fang_local_2020,baruch_little_2019}. We use $m$ to denote the number of malicious clients. The batch size $|B|$ is set to range from 8 to 32. More detailed settings can be found in~\ref{appendix:details}.

\paragraph{Evaluated AGRs} 
We consider the following four representative Byzantine-robust AGRs in our experiments.
\begin{itemize}
    \item \textbf{Multi-krum.}  \textit{Multi-krum}~\cite{blanchard_machine_2017} is based on the intuition that the malicious gradients should be far from the benign gradients in the gradient space. It compares the pair-wise distance of local updates from clients, and selects the top $c$ (where $N-c>2m+2$) trustworthy updates.
    \item \textbf{Bulyan.} \textit{Bulyan}~\cite{mhamdi_hidden_2018} is also based on distance to remove malicious gradients. It first selects $c_1$ gradients where $N-c_1\geq2m$. Then it calculates the median gradient based on the dimensions of the selected gradient set, and averages the $c_2$ gradients that are closest to the median gradient where $c_1-c_2\geq2m$. 
    It requires $N\geq 4m+3$.
    \item \textbf{Adaptive federated average (AFA).} \textit{AFA}~\cite{cao_fltrust_2022} 
    introduces the state-of-the-art trust bootstrapping mechanism, which leverages a small and clean dataset to detect and then drop the suspiciously malicious local updates of which the cosine similarity compared to the reference gradient is less than a threshold. 
    \item \textbf{Fang defenses.} \textit{Fang defenses}~\cite{fang_local_2020} is another state-of-the-art AGR based on the rejection of loss and rejection of error. It detects and drops anomalous local updates based on their impact on the prediction error rate and the loss. 
\end{itemize}
{}

\subsection{Baselines and Evaluation Metrics}
\paragraph{Baselines} We compare the \attack with the following baselines. They are the state-of-the-art model/gradient inversion attacks that can be launched from the client side in FL. 
\begin{itemize}
    \item \textbf{DLG.} \textit{DLG}~\cite{zhu_deep_2019} is the beginning work in gradient inversion attack. It tries to optimize random noise and random labels to fit the target gradient based on $l_2$ distance.
    \item \textbf{iDLG.} \textit{iDLG}~\cite{zhao_idlg_2020} improves \textit{DLG} by inferring true labels analytically. Generally, \textit{iDLG} is better than \textit{DLG} on inversed image quality and inversion attack success probability.
    \item \textbf{IG.} \textit{IG}~\cite{geiping_inverting_2020} mainly uses cosine distance to optimize random noise to approach the target gradient. This work emphasized and validated that gradient direction is more important.
\end{itemize}
It should be noted that we do not select \textit{GradInversion}~\cite{yin_see_2021} as one of baselines as it requires batch-wise mean and variance, which are impossible to obtain for a client-side adversary. 
In addition, given that \textit{iDLG} and \textit{IG} lack effective label inference methods from the client side, we provide these two baselines with label information. Such experimental setting give the baseline attacks additional advantage. 



\paragraph{Evaluation metrics} 
We use three following metrics: i) The peak signal-to-noise ratio (PSNR). It compares the original and the reconstructed image or video to quantify the amount of distortion. 
ii) The learned perceptual image patch similarity (LPIPS)~\cite{zhang2018unreasonable}. LPIPS is a distance metric for image quality assessment and aims to measure the similarity between the two images based on distance in the embedding space of the VGG network. iii) The structural similarity index measure (SSIM)~\cite{wang2004image}. SSIM takes into account the changes in structural information, luminance, and contrast that occur between the two images and measures the similarity between them. The higher values of PSNR and SSIM, and the lower value of LPIPS indicate better reconstruction.

\section{Evaluation Results}
\subsection{\attack Performance} 
\begin{figure}[H]
	\centering
	\subfloat[Cifar100.]{\includegraphics[width=0.45\textwidth]{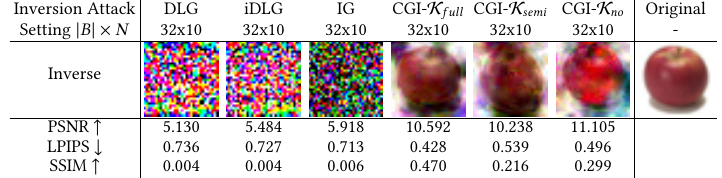}%
		\label{subfig:CompGradInversionCifar100}}
	\hfil
	\subfloat[TinyImageNet.]{\includegraphics[width=0.45\textwidth]{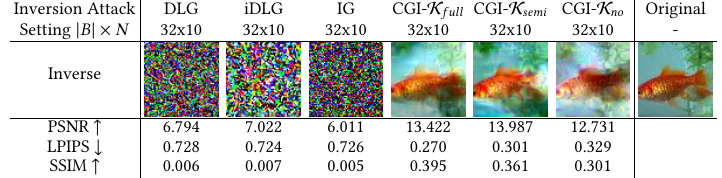}%
		\label{subfig:CompGradInversionTinyImageNet}}
    \hfil
	\subfloat[CalTech256.]{\includegraphics[width=0.45\textwidth]{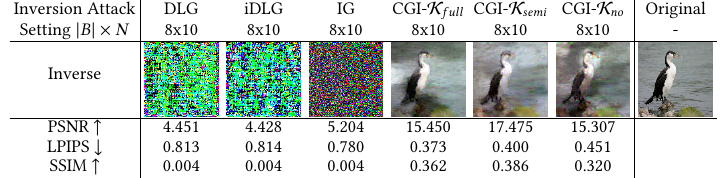}%
		\label{subfig:CompGradInversionCalTech256}}
	\caption{Comparison of \attack with baseline methods.}
	\label{fig:CompGradInversion}
\end{figure}

\paragraph{Baseline performance} 
Figure~\ref{fig:CompGradInversion} shows the comparison of our \attack with DLG~\cite{zhu_deep_2019}, iDLG~\cite{zhao_idlg_2020} and IG~\cite{geiping_inverting_2020}. 
We do not equip any Byzantine-robust AGRs in this set of experiments for a fair comparison with the baselines. 
The results demonstrate that only \attack can invert target samples successfully as a client-side adversary when the batch size is set to 32 on CIFAR100 and TinyImageNet and 8 on CalTech256. 

\begin{figure*}[!htp]
	\centering
	\subfloat[\attack on Cifar100]{\includegraphics[width=1\textwidth]{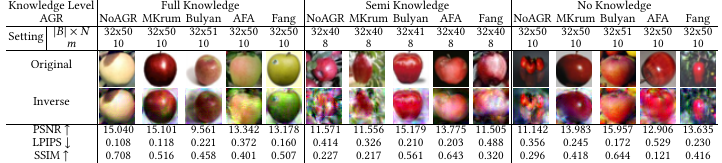}%
		\label{Fig_InversionResultSubCifar}}
	\hfil
	\subfloat[\attack on TinyImageNet]{\includegraphics[width=1\textwidth]{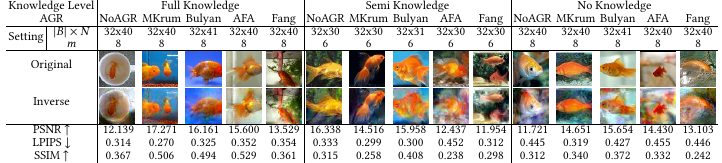}%
		\label{Fig_InversionResultSubImageNet}}
    \hfil
	\subfloat[\attack on CalTech256 ]{\includegraphics[width=1\textwidth]{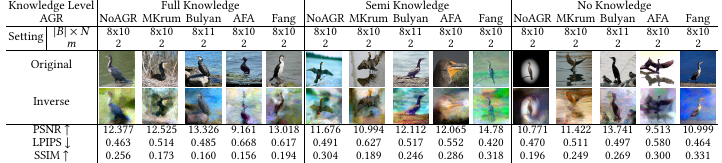}%
		\label{Fig_InversionResultSubCalTech}}
	\caption{\attack's performance against AGRs.}
	\label{fig:InversionResult}
\end{figure*}

\paragraph{\attack against Byzantine-robust AGRs} 
{}
The \attack's performance against AGRs is illustrated in Figure~\ref{fig:InversionResult}. In general, \attack is capable of defeating all evaluated robust AGRs across various datasets. 
While some restored images may contain noise, the essential semantic features of the targeted images can still be revealed through our attack. 
The reconstructed image quality is not highly sensitive to the choice of robust AGRs as we consistently observe PSNR values exceeding 10.00, LPIPS scores below 0.60, and SSIM values above 0.2 across all settings and datasets. 
For example, our \attack achieves LIPIS($\downarrow$)  values of 0.377, 0.358, 0.350, 0.462, and 0.387 against \textit{NoAGR}, \textit{Multi-krum}, \textit{Bulyan}, \textit{AFA}, and \textit{Fang}, respectively, across all knowledge levels and datasets in average. 
We observe that the LPIPS value on AFA is slightly higher, while the LPIPS values on other robust AGRs are similar. 
This can be explained by the success of our attack relying on the targeted gradient's ability to dominate the aggregated gradient. 
Since AFA uses a weighted average aggregation method, the influence of the targeted gradient is slightly impaired during aggregation, resulting in slightly worse performance compared to other AGRs. Generally, we only consider just one targeted class image joining the aggregation. The study of more targeted samples in the aggregation can be found in~\ref{appendix:targeted_distribution}

\subsection{Performance of Break-down Attack Steps}\label{sec:breakdownsteps} 
In this section, we evaluate the performance of \attack in crafting the malicious model and poisoning the global model (i.e., the first two steps) and investigate their influence.  

\paragraph{Performance of the malicious model crafting} 
This step aims to craft a malicious model that  maximizes the loss
of the targeted class while performing normally on the main task, we use the accuracy and  cross-entropy loss to evaluate them. 


Table~\ref{table:MaliciousModelPerformance} illustrates the results. 
We find that all adversaries achieve $0.00\%$ of accuracy with respect to the targeted class while retaining performance on the main task basically. However, the adversary with full knowledge achieves a larger loss gap between the targeted class and the main task than the adversaries with semi/no knowledge. This is because the adversary can control targeted class loss with full knowledge and generally make target class loss very large. 
The adversary can take advantage of such a gap toward a more effective attack. In addition, the adversary with different knowledge levels all retains the main task performance. This proves that \attack is a sneaky attack.

\begin{table}[!htbp]
\centering
\caption{Evaluation of Malicious Models}
\label{table:MaliciousModelPerformance}
\resizebox{\linewidth}{!}{
\begin{tblr}{
  cells = {c},
  cell{1}{1} = {r=2}{},
  cell{1}{2} = {r=2}{},
  cell{1}{3} = {c=3}{},
  cell{3}{1} = {r=4}{},
  cell{7}{1} = {r=4}{},
  cell{11}{1} = {r=4}{},
  hline{1,3,7,11,15} = {-}{},
  hline{5,9,13} = {2-5}{},
  hline{2} = {3-5}{},
}
{\textbf{Adversary’s} \\ \textbf{Knowledge}}  & \textbf{Metric}            & \textbf{Datasets} &              &            \\
                            &                   & Cifar100 & TinyImageNet & CalTech256 \\
{Full \\ knowledge}          & Targeted Class Acc.(\%) & 0.000   & 0.000   & 0.000   \\
                            & Targeted Class Loss     & 262.382 & 310.962 & 108.678 \\
                            & Main Task Acc.(\%)    & 96.301  & 97.685  & 96.468  \\
                            & Main Task Loss        & 0.153   & 0.162   & 0.156   \\
{Semi \\ knowledge}        & Targeted Class Acc.(\%) & 0.000   & 0.000   & 0.000   \\
                            & Targeted Class Loss     & 30.581  & 28.584  & 20.574  \\
                            & Main Task Acc.(\%)    & 97.929  & 99.715  & 99.793  \\
                            & Main Task Loss        & 0.070   & 0.009   & 0.016   \\
{No \\ knowledge}        & Targeted Class Acc.(\%) & 0.000   & 0.000   & 0.000   \\
                            & Targeted Class Loss     & 35.355  & 47.100  & 67.236  \\
                            & Main Task Acc.(\%)    & 99.956  & 99.345  & 99.497  \\
                            & Main Task Loss        & 0.002   & 0.021   & 0.030  
\end{tblr}
}
\vspace{+1em}
\end{table}

\paragraph{Performance of the global model poisoning} 
In this set of evaluation,  we focus on the poisoned global model's performance with respect to the targeted class and the main task on the training dataset. 
We regard the distances between the global models and malicious models before poisoning as the initial distance and after poisoning as the final distance, which is calculated as $\sqrt{\sum^l_{l_0=1}(\phi_G-\phi_{mal})^2/d_\phi}$ where $l$ represents the number of layers 
and $d_\phi$ is the parameter dimension. When the final replacement distance is approaching zero, we can regard the poisoning  processing as successful.

We give the result in Table~\ref{table:ModelRep}. Our model poisoning  succeeds in all settings.  Almost all global models' performance on accuracy and loss are close to the corresponding malicious models (as presented in Table~\ref{table:MaliciousModelPerformance}). Among the three knowledge levels, the adversary with semi-knowledge achieves the smallest initialized model distance ($61.046 \times 10^{-3}$, $89.778 \times 10^{-3}$ and $40.019 \times 10^{-3}$ on Cifar100, TinyImageNet and CalTech256 respectively), followed by the adversary with no-knowledge ($1087.533 \times 10^{-3}$, $128.167 \times 10^{-3}$ and $217.541 \times 10^{-3}$ on Cifar100, TinyImageNet and CalTech256 respectively). The full-knowledge adversary has the largest initialized model distance ($2157.039 \times 10^{-3}$, $5082.850 \times 10^{-3}$ and $288.970 \times 10^{-3}$ on Cifar100, TinyImageNet and CalTech256 respectively), meaning that it requires more poisoning iterations than the other two to make the final poisoned model close to the malicious model. 
Comparing different AGRs, we find that Fang and AFA are more robust as they achieve lower loss with respect to the targeted class than other AGRs,  
and the final distances of AFA and Fang are also generally more distinct than other AGRs.
This is due to the use of the validation dataset. Particularly, in the extreme case that the validation dataset only consists of samples from the target class, our poisoning will fail. 

The effectiveness of gradient inversion in \attack is influenced by both the performance of malicious model crafting and global model poisoning in practical environments. The malicious models crafted in step 1 serve as the foundation for \attack, as the loss gap between the targeted class and the main task determines whether the targeted gradient can dominate the aggregated gradient. 
For poisoning the global model (step 2), different AGR determines the difficulty of model replacement, which determines whether the malicious updates can be successfully injected into the global model.

\DefTblrTemplate{caption}{default}{}
\begin{table}[!t]
\Huge
\caption{The performance of the global model poisoning.}
\label{table:ModelRep}
\resizebox{1\linewidth}{!}{
\begin{talltblr}[
    note{1}={"T.C." stands for "Targeted Class".},
    note{2}={"M.T." stands for "Main Task".},
    note{3}={"I.M.Dis." means "Initialized Model Distance".},
    note{4}={"F.M.Dis." means "Final Model Distance".}
]{
  width=0.5\textwidth,
  cells = {c,m},
  cell{1}{1} = {r=2}{},
  cell{1}{2} = {c=2}{},
  cell{1}{4} = {c=6}{},
  cell{3}{1} = {c=3}{},
  cell{9}{1} = {c=3}{},
  cell{15}{1} = {c=3}{},
  cell{21}{1} = {c=3}{},
  cell{27}{1} = {c=3}{},
  cell{33}{1} = {c=3}{},
  cell{39}{1} = {c=3}{},
  cell{45}{1} = {c=3}{},
  cell{51}{1} = {c=3}{},
  vline{2,3} = {-}{},
  vline{1,4,6,8,10} = {1-2,4-8,10-14,16-20,22-26,28-32,34-38,40-44,46-50,52-56}{},
  hline{1,2,3} = {-}{},
  hline{4,9,10,15,16,21,22,27,28,33,34,39,40,45,46,51,52,57} = {-}{}
}
AGR           & Setting                    &    & Metric    &           &           &           &            &            \\
    & {$|B|\times N$} &{$m$}                                & {T.C.\TblrNote{1} \\ Acc. (\%)} & {T.C.\TblrNote{1} \\ Loss} & {M.T.\TblrNote{2} \\ Acc. (\%)} & {M.T.\TblrNote{2} \\ Loss} & {I.M.\\Dis.\TblrNote{3} ($10^{-3}$)} & {F.M.\\ Dis.\TblrNote{4} ($10^{-3}$)} \\
Cifar100 - $\mathcal{K}_{full}$      &                            &    &           &           &           &           &            &            \\
No Robust AGR & 32x50                      & 10 & 0.000     & 229.426   & 96.319    & 0.153     & 2157.039   & 3.203      \\
Multi-Krum    & 32x50                      & 10 & 0.000     & 263.485   & 96.317    & 0.153     & 2157.039   & 0.446      \\
Bulyan        & 32x51                      & 10 & 0.000     & 263.809   & 96.305    & 0.153     & 2157.039   & 0.039      \\
AFA           & 32x50                      & 10 & 0.000     & 222.714   & 96.889    & 0.124     & 2157.039   & 318.045    \\
Fang          & 32x50                      & 10 & 0.000     & 185.995   & 96.348    & 0.153     & 2157.039   & 5.944      \\
Cifar100 - $\mathcal{K}_{semi}$      &                            &    &           &           &           &           &            &            \\
No Robust AGR & 32x40                      & 8  & 0.000     & 27.679    & 97.497    & 0.083     & 61.046     & 1.798      \\
Multi-Krum    & 32x40                      & 8  & 0.000     & 30.377    & 96.497    & 0.108     & 61.046     & 0.473      \\
Bulyan        & 32x41                      & 8  & 0.000     & 30.431    & 94.764    & 0.155     & 61.046     & 0.254      \\
AFA           & 32x40                      & 8  & 0.000     & 18.815    & 98.667    & 0.052     & 61.046     & 17.086     \\
Fang          & 32x40                      & 8  & 0.000     & 27.401    & 95.273    & 0.142     & 61.046     & 4.437      \\
Cifar100 - $\mathcal{K}_{no}$      &                            &    &           &           &           &           &            &            \\
No Robust AGR & 32x50                      & 10 & 0.000     & 30.525    & 99.978    & 0.001     & 1087.533   & 2.823      \\
Multi-Krum    & 32x50                      & 10 & 0.000     & 34.375    & 99.976    & 0.001     & 1087.533   & 0.018      \\
Bulyan        & 32x51                      & 10 & 0.000     & 29.263    & 99.980    & 0.001     & 1087.533   & 45.807     \\
AFA           & 32x50                      & 10 & 0.000     & 30.887    & 99.978    & 0.001     & 1087.533   & 2.427      \\
Fang          & 32x50                      & 10 & 0.000     & 30.221    & 99.978    & 0.001     & 1087.533   & 2.958      \\
TinyImageNet - $\mathcal{K}_{full}$  &                            &    &           &           &           &           &            &            \\
No Robust AGR & 32x40                      & 8  & 0.000     & 307.947   & 97.676    & 0.164     & 5082.850   & 1.800      \\
Multi-Krum    & 32x40                      & 8  & 0.000     & 317.606   & 97.689    & 0.165     & 5082.850   & 0.091      \\
Bulyan        & 32x41                      & 8  & 0.000     & 316.568   & 97.700    & 0.163     & 5082.850   & 23.094     \\
AFA           & 32x40                      & 8  & 0.000     & 317.495   & 97.370    & 0.186     & 5082.850   & 39.341     \\
Fang          & 32x40                      & 8  & 0.000     & 211.890   & 97.712    & 0.162     & 5082.850   & 14.497     \\
TinyImageNet - $\mathcal{K}_{semi}$  &                            &    &           &           &           &           &            &            \\
No Robust AGR & 32x30                      & 6  & 0.000     & 27.227    & 99.834    & 0.006     & 89.778     & 1.025      \\
Multi-Krum    & 32x30                      & 6  & 0.000     & 28.284    & 99.784    & 0.007     & 89.778     & 0.078      \\
Bulyan        & 32x31                      & 6  & 0.000     & 27.789    & 99.832    & 0.006     & 89.778     & 2.044      \\
AFA           & 32x30                      & 6  & 0.000     & 27.024    & 99.822    & 0.006     & 89.778     & 0.625      \\
Fang          & 32x30                      & 6  & 0.000     & 28.174    & 99.809    & 0.007     & 89.778     & 0.544      \\
TinyImageNet - $\mathcal{K}_{no}$  &                            &    &           &           &           &           &            &            \\
No Robust AGR & 32x40                      & 8  & 0.000     & 44.799    & 99.759    & 0.009     & 128.167    & 1.367      \\
Multi-Krum    & 32x40                      & 8  & 0.000     & 45.790    & 99.649    & 0.012     & 128.167    & 0.263      \\
Bulyan        & 32x41                      & 8  & 0.000     & 45.177    & 99.829    & 0.006     & 128.167    & 1.878      \\
AFA           & 32x40                      & 8  & 0.000     & 36.747    & 99.959    & 0.002     & 128.167    & 21.990     \\
Fang          & 32x40                      & 8  & 0.000     & 41.507    & 99.866    & 0.006     & 128.167    & 2.622      \\
CalTech256 - $\mathcal{K}_{full}$    &                            &    &           &           &           &           &            &            \\
No Robust AGR & 8x10                       & 2  & 0.000     & 114.571   & 96.413    & 0.164     & 288.970    & 2.448      \\
Multi-Krum    & 8x10                       & 2  & 0.000     & 136.099   & 96.167    & 0.198     & 288.970    & 43.910     \\
Bulyan        & 8x11                       & 2  & 0.000     & 118.328   & 96.388    & 0.163     & 288.970    & 9.523      \\
AFA           & 8x10                       & 2  & 0.000     & 94.622    & 64.082    & 1.533     & 288.970    & 195.884    \\
Fang          & 8x10                       & 2  & 0.000     & 39.601    & 91.493    & 0.351     & 288.970    & 174.439    \\
CalTech256 - $\mathcal{K}_{semi}$    &                            &    &           &           &           &           &            &            \\
No Robust AGR & 8x10                       & 2  & 0.000     & 20.364    & 99.831    & 0.011     & 40.019     & 4.474      \\
Multi-Krum    & 8x10                       & 2  & 0.000     & 20.528    & 99.814    & 0.014     & 40.019     & 0.603      \\
Bulyan        & 8x11                       & 2  & 0.000     & 20.561    & 99.805    & 0.015     & 40.019     & 0.857      \\
AFA           & 8x10                       & 2  & 0.000     & 25.588    & 99.023    & 0.042     & 40.019     & 11.968     \\
Fang          & 8x10                       & 2  & 0.000     & 20.377    & 99.822    & 0.014     & 40.019     & 1.603      \\
CalTech256 - $\mathcal{K}_{no}$    &                            &    &           &           &           &           &            &            \\
No Robust AGR & 8x10                       & 2  & 0.000     & 70.446    & 99.797    & 0.013     & 217.541    & 6.819      \\
Multi-Krum    & 8x10                       & 2  & 0.000     & 67.188    & 99.539    & 0.027     & 217.541    & 0.718      \\
Bulyan        & 8x11                       & 2  & 0.000     & 70.383    & 99.704    & 0.020     & 217.541    & 9.314      \\
AFA           & 8x10                       & 2  & 0.000     & 71.169    & 99.200    & 0.039     & 217.541    & 11.477     \\
Fang          & 8x10                       & 2  & 0.000     & 48.784    & 99.894    & 0.008     & 217.541    & 22.017     
\end{talltblr}
}
\vspace{+1em}
\end{table}


\subsection{Ablation Studies}\label{sec:ablation}
\paragraph{Effect of the knowledge of data distribution} 
We first investigate how different levels of the adversary's knowledge about the data distribution affect the attack's performance.

\begin{figure}[t]
	\centering
	\subfloat[Cifar100]{\includegraphics[width=0.5\textwidth]{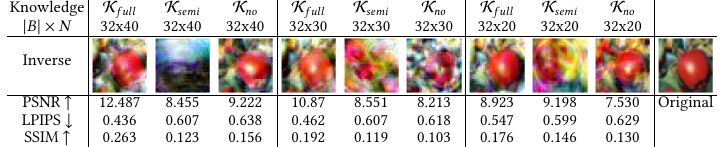}%
		\label{Fig_ModelComparisonSubCifar100}}
	\hfil
	\subfloat[TinyImageNet]{\includegraphics[width=0.5\textwidth]{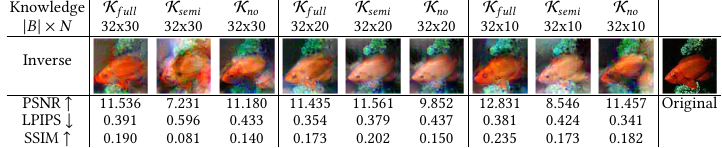}%
		\label{Fig_ModelComparisonSubTinyImageNet}}
    \hfil
	\subfloat[CalTech256]{\includegraphics[width=0.5\textwidth]{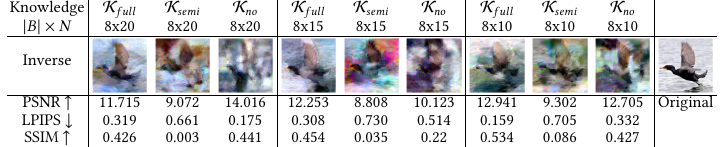}%
		\label{Fig_ModelComparisonSubCalTech256}}
	\caption{\attack's performance with different levels of knowledge.}
	\label{fig:ModelComparison}
    \vspace{+1em}
\end{figure}

We fix the batch size in each set of comparisons among the three knowledge levels. 
As shown in Figure~\ref{fig:ModelComparison}, the full knowledge adversary achieves the best performance, and its performance remains stable with the increasing number of clients. The adversary with semi-knowledge and no-knowledge's averaged performance are similar, while being unstable in some cases (e.g., Cifar100 inversion results under 20 and 30 clients).  
Basically, the full knowledge adversary has a larger loss gap between non-targeted and targeted class samples, which has been illustrated in Table~\ref{table:MaliciousModelPerformance}. Because of the larger loss gap, the targeted gradient on the malicious model trained with full knowledge is hard to be diluted during aggregation. 
Therefore, full knowledge malicious model can work on larger FL settings. 
In comparison to a full-knowledge adversary, semi-knowledge and no-knowledge methods utilize targeted class knowledge to poison models. However, the limited knowledge of the targeted class can pose challenges. This limitation can result in scenarios where the targeted class loss is not significant enough which may impair inversion performance, or the main task loss increases due to the potential damage to generalization features. 
Hence, it can be observed that malicious models crafted with semi-knowledge or no knowledge are prone to instability in certain cases.   



\paragraph{Effect of the batch size and the number of clients} 
In \attack, the effect of the injected maliciousness with respect to the targeted sample is affected by the batch size and the number of clients {\iffalse as the sum of the local updates will be divided by the sample size in one aggregation (i.e., $|B| \times N$) to form the aggregated updates according to the rule of gradient descent. \fi} as the sum of the local updates will gradually dominate the aggregated gradient and dilute the targeted gradient when the number of training samples increases in one agggregation.
We thus conduct the following experiments to explore the impact. 

\begin{figure*}[t]
	\centering
	\subfloat[PSNR]{\includegraphics[width=2in]{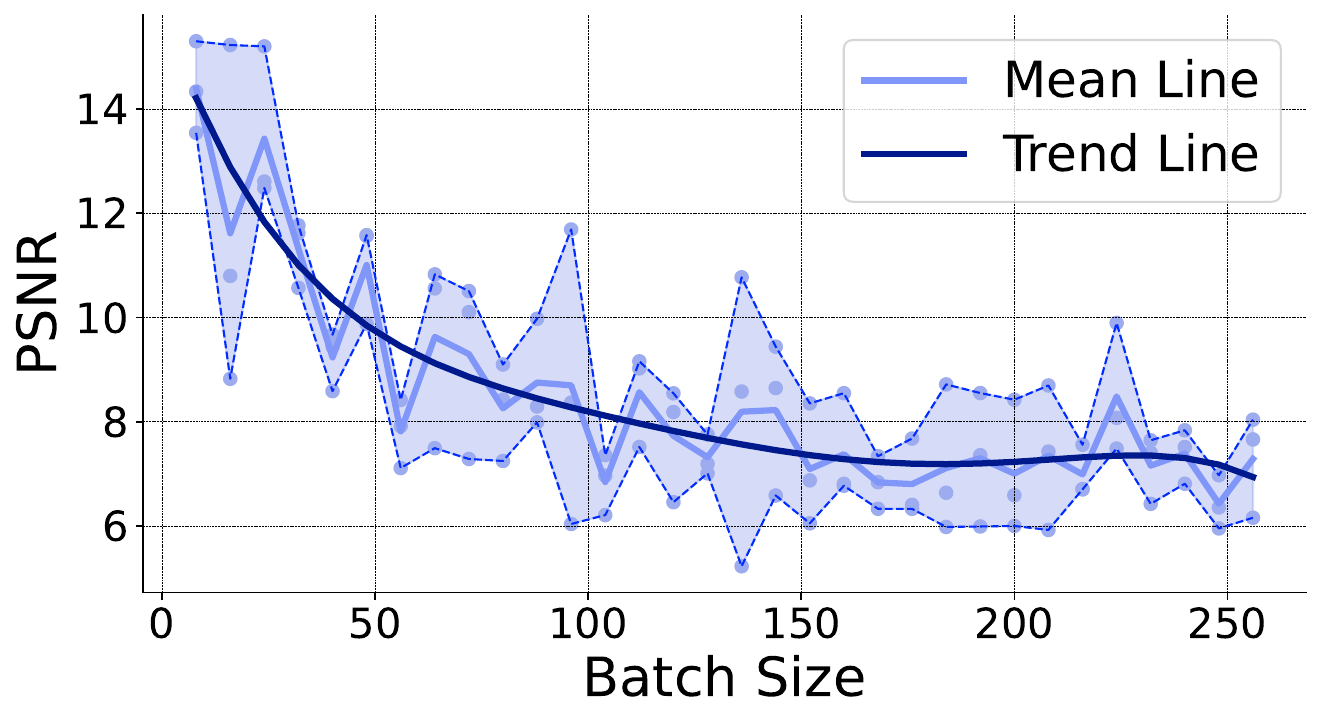}%
		\label{Fig_BatchEffectSubPSNR}}
	\hfil
	\subfloat[LPIPS]{\includegraphics[width=2in]{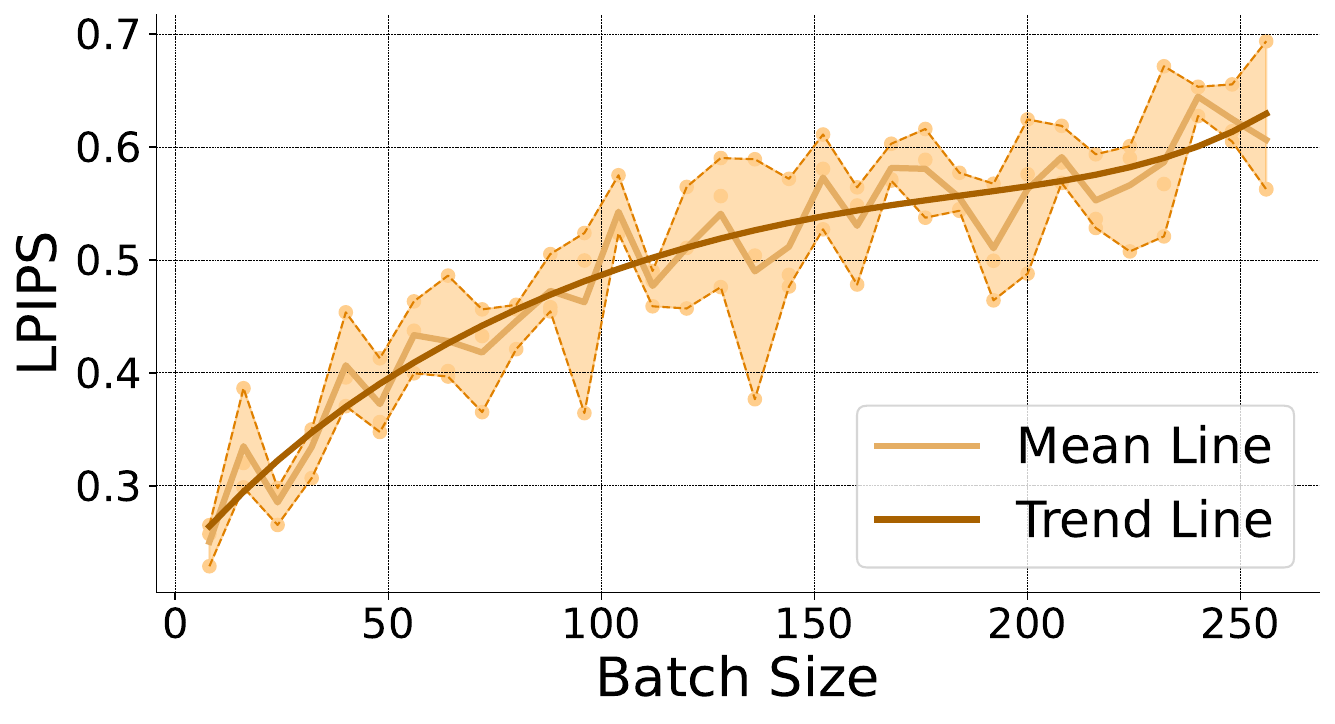}%
		\label{Fig_BatchEffectSubLPIPS}}
    \hfil
	\subfloat[SSIM ]{\includegraphics[width=2in]{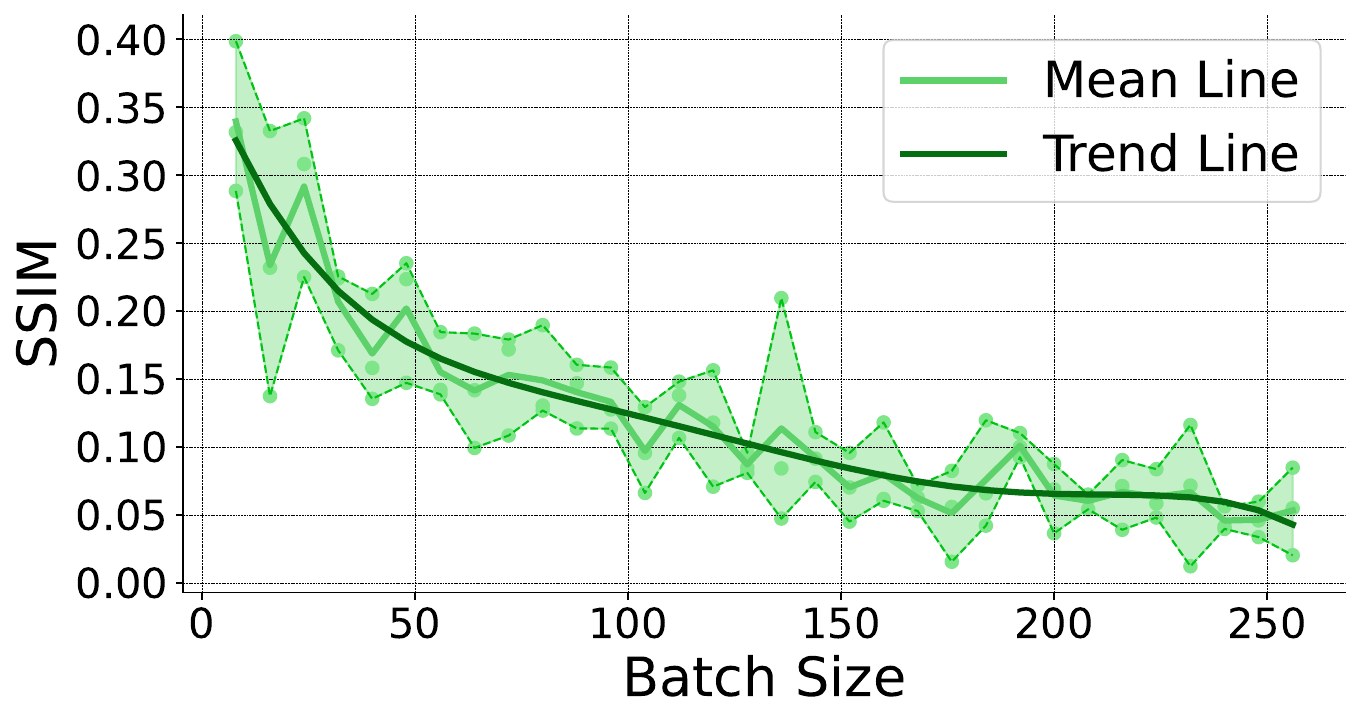}%
		\label{Fig_BatchEffectSubSSIM}}
	\caption{PSNR, LPIPS, and SSIM of \attack against the batch size in one aggregation of 50 clients on TinyImageNet.}
	\label{fig:BatchEffectChart}
\end{figure*}

Figure~\ref{fig:BatchEffectChart} shows the \attack's performance with the varying number of batch sizes in one aggregation of 50 clients. The experimental setting is fixed with the full-knowledge adversary without any robust AGRs equipped. We can find that the performance is going down as the batch size increases.  
It is worth noting that in this set of experiments, the level of maliciousness is kept fixed, resulting in a fixed loss gap. However, an adaptive attacker can actively increase the level of maliciousness in order to enlarge the loss gap, especially for larger batch sizes. By doing so, the attacker can maintain a high level of attack performance.



We also conduct experiments to fix the batch size and test on various numbers of clients in~\ref{appendix:number_client}, and the performance trend is similar.

\begin{figure*}[]
	\centering
	\subfloat[PSNR]{\includegraphics[width=2in]{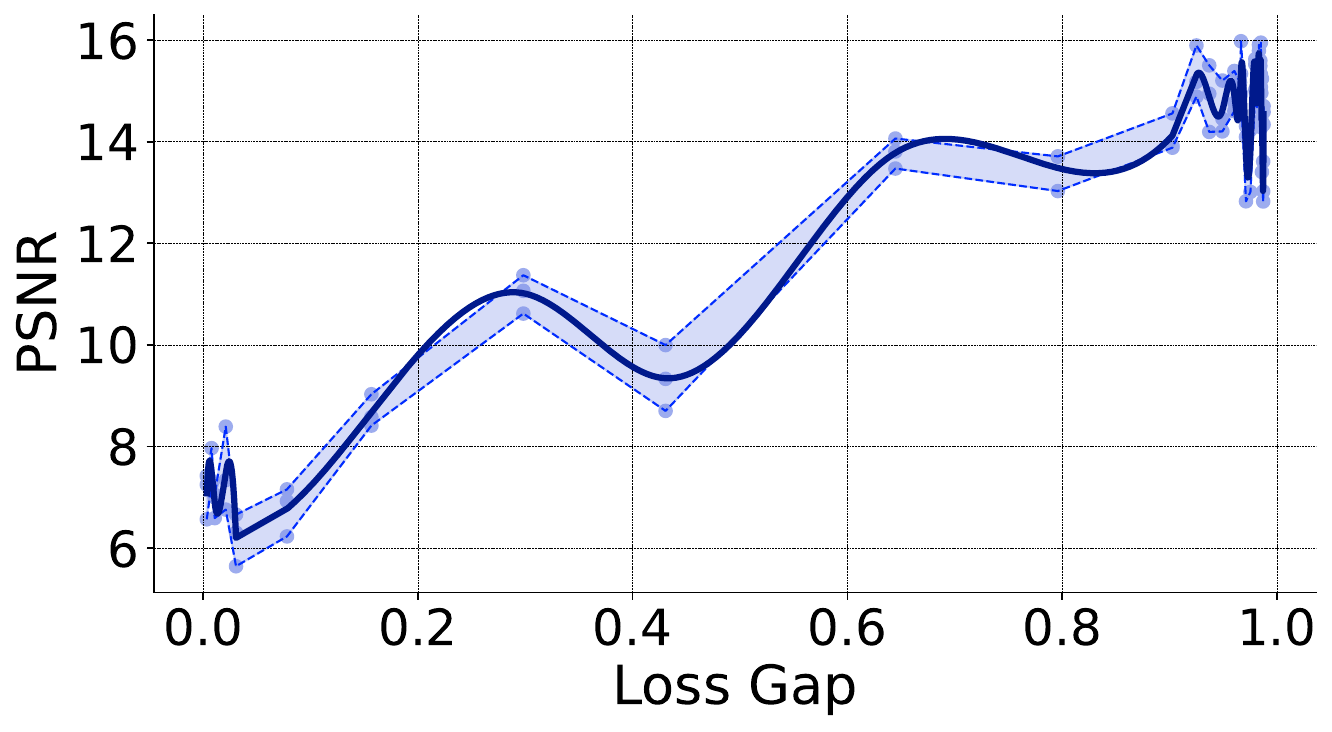}%
		\label{Fig_LossGapSubPSNR}}
	\hfil
	\subfloat[LPIPS]{\includegraphics[width=2in]{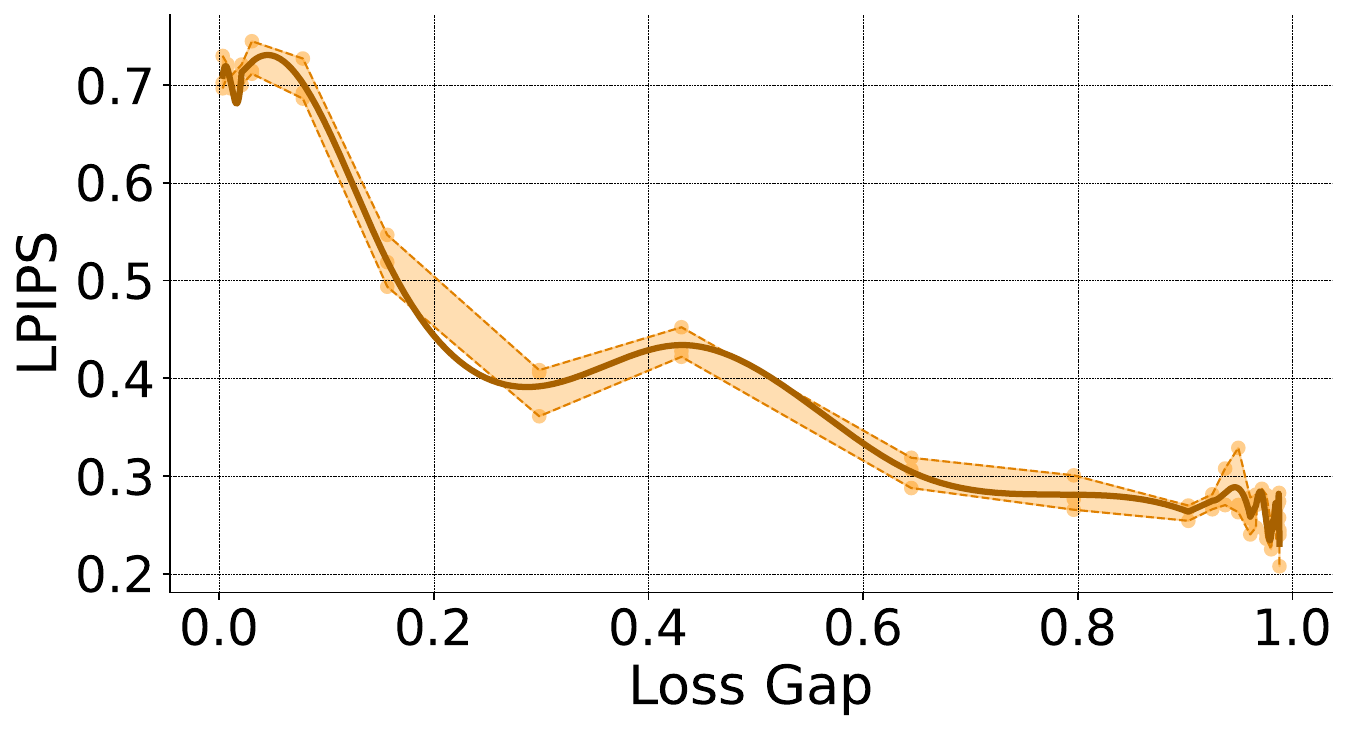}%
		\label{Fig_LossGapSubLPIPS}}
    \hfil
	\subfloat[SSIM]{\includegraphics[width=2in]{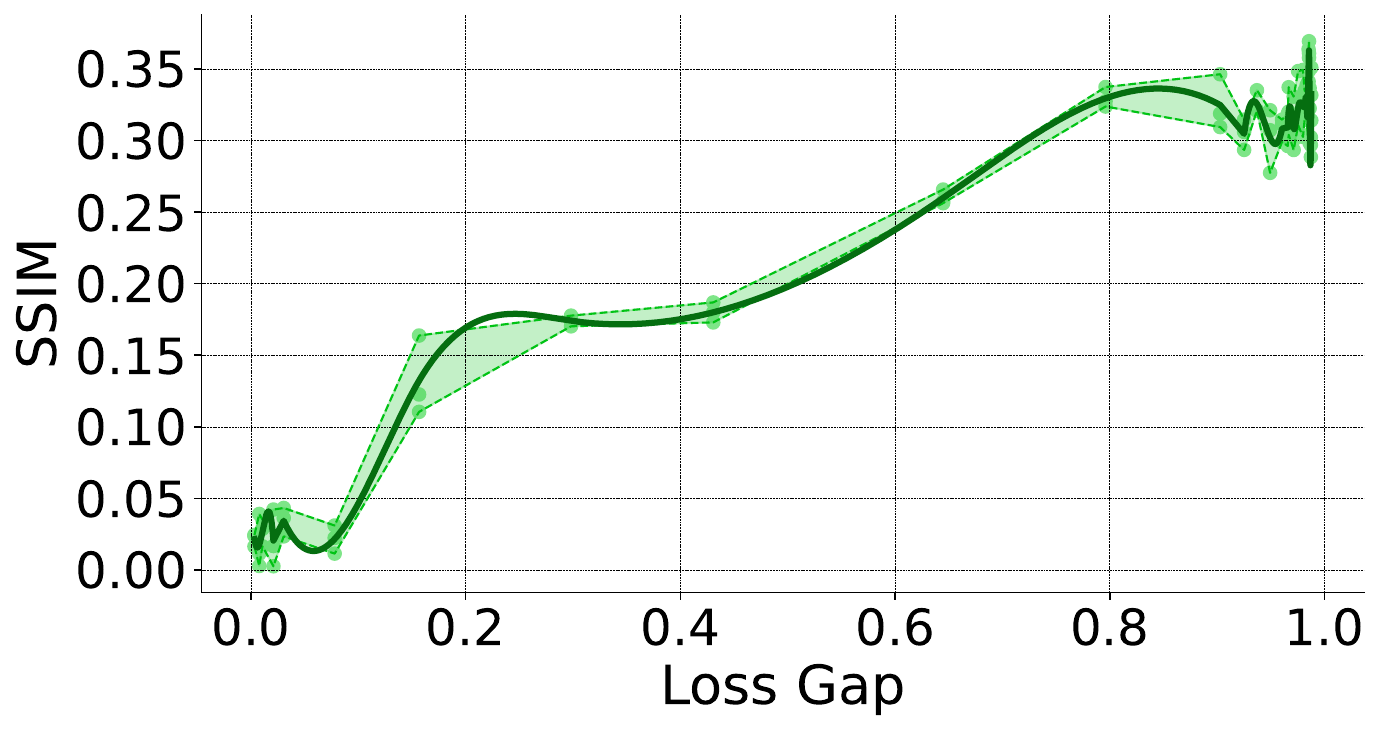}%
		\label{Fig_LossGapSubSSIM}}
	\caption{\attack's performance on TinyImageNet against the varying loss gap.}
	\label{fig:LossGapChart}
\end{figure*}

\paragraph{Effect of the loss gap} 
This set of experiments explores the impact of the gap between the loss of the targeted samples and the loss of the main task on the performance of \attack.   

{}

We denote the loss gap as
\begin{equation}
    \mathcal{L}_{gap} = \frac{\Delta\mathcal{L}_{tar}}{\Delta\mathcal{L}_{tar}+\Delta\mathcal{L}_{main}},
\end{equation}

where $\Delta\mathcal{L}_{tar} = |\mathcal{L}(\Phi_G^{k+1},X_{tar})-\mathcal{L}(\Phi_G^{k},X_{tar})|$,
representing the loss variation of the targeted samples with respect to the poisoned global model $\Phi_G^t$ and $\Phi_G^{t+1}$, and 
$\Delta\mathcal{L}_{main} = |\mathcal{L}(\Phi_G^{k+1},X_{main})-\mathcal{L}(\Phi_G^{k},X_{main})|, $
representing the loss variation of the remaining samples with respect to the poisoned global model $\Phi_G^k$ and $\Phi_G^{k+1}$. 
As such, $\mathcal{L}_{gap}$ is between 0 and 1, and the more the overall loss is dominated by $\Delta\mathcal{L}_{tar}$, the closer $\mathcal{L}_{gap}$ is to 1. 


Our experiments are conducted on TinyImageNet~\cite{tiny-imagenet} with 40 clients and batch size of 32. The adversary is of full knowledge.  
Figure~\ref{fig:LossGapChart} demonstrates how the loss gap affects the \attack's performance. When the $\mathcal{L}_{gap}$ is small,  the targeted gradient is  mixed with other samples' features together, resulting in the low performance of \attack. 
With the increase of the $\mathcal{L}_{gap}$, the \attack's performance increases as the targeted loss becomes distinct. When $\mathcal{L}_{gap}$ approaches 1, which means the gradient almost only contains the targeted gradient, the best inversion performance is achieved under this condition. Among different metrics, PSNR has a larger variance because the randomness in inversion attack may change the background of reconstructed images. LPIPS and SSIM basically keep stable.

\section{Defenses and Mitigation Strategies}

 \attack represents a novel client-side attack in FL. Consequently, defending against this type of attack remains an open topic. We analyze existing privacy-preserving methods and their effectiveness in mitigating our proposed attack. Additionally, we propose potential directions for developing defense mechanisms.

 Firstly, secure aggregation based on Secure Multi-Party Computation (MPC) solutions~\cite{BonawitzMPC2017} is not effective in defending against our attack. Although secure aggregation can prevent local updates from exposure, the aggregated update remains disclosed. In \attack, once the global model has been poisoned, the targeted gradient can be exposed through the aggregated update. 
 Secondly, Differential Privacy (DP)~\cite{abadi2016deep} can be ineffective in defending our attack as well. DP relies on the indistinguishability of outputs when inputs are any two individual samples in the dataset.
 Such privacy promise holds in the FL when all participants follow the learning protocol, as the local gradients (which determines the sensitivity of the algorithm) can be clipped and unified. However, a poisoning attack can simply break the DP privacy guarantee as the feeding of the malicious updates can largely affect the output space of the algorithm. 
 In \attack, the global model is poisoned by significantly amplifying the targeted gradient. Consequently, a substantially larger amount of noise is required to maintain the same level of privacy based on the maliciously amplified sensitivity. However, excessively adding noise to the update would undermine the generalization capability of the global model. 
 Thirdly,  defense methods based on de-duplicating training data (i.e., reducing the repeated use of the data samples)~\cite{kandpalDeduplicating22} are incapable of mitigating our attack since our attack does not rely on  memorization. Even if a targeted sample appears only once in the training, it can still be reconstructed by \attack.

 For defending against \attack, we suggest potential countermeasures that aim at mitigating the poisoning effect. 
 Since the attack specifically targets the private training samples of honest clients, incorporating defense mechanisms involving benign clients can be effective in thwarting the attack.
 One approach is to have benign clients calculate the accuracy and loss of their training data in each round. If certain samples show a significant increase in loss, the benign client can identify them as potentially compromised and take actions such as discarding those samples. 
 However, it is crucial to carefully consider and address the potential impact of homogeneity among the data samples. Removing too many samples that are similar to the targeted sample may adversely affect the model's generalization ability on the targeted class. 
 Additionally, employing a server-side Byzantine-robust with a stronger assumption on the prior knowledge of the entire data distribution can be an effective defense against our attack. 
 For examples, the robust AGRs~\cite{fang_local_2020, cao_fltrust_2022} that leverage a validation dataset to detect the malicious updates can be enhanced by including a validation dataset that contains sufficient samples from each class. 


\section{Conclusion}
We have introduced \attack to invert privacy image data from  malicious clients. \attack works as a novel and stealthy client-side attack, which allows the adversary who can only obtain the least knowledge from FL to reconstruct training samples from other clients. 
One notable aspect of \attack is its capability to invert gradients from a batch. 
Moreover, it has demonstrated its effectiveness in defeating Byzantine-robust AGRs, which are designed to withstand poisoning attacks in FL settings. 
Through extensive experiments, we have substantiated the effectiveness and stealthiness of \attack, highlighting its potential to compromise privacy and the need for further development of dedicated robust defense mechanisms specifically designed to mitigate the threat posed by \attack. 


In future work, we will further study the underlying mechanism of the inversion attack and how gradients carry original data features. Additionally, we aim to explore and develop effective strategies, specifically tailored to mitigate the risks associated with inversion attacks against FL.

\bibliographystyle{plain}
\bibliography{reference}

\begin{thebibliography}{10}

\bibitem{abadi2016deep}
Martin Abadi, Andy Chu, Ian Goodfellow, H~Brendan McMahan, Ilya Mironov, Kunal
  Talwar, and Li~Zhang.
\newblock Deep learning with differential privacy.
\newblock In {\em Proceedings of the 2016 ACM SIGSAC conference on computer and
  communications security}, pages 308--318, 2016.

\bibitem{bagdasaryan_how_2020}
Eugene Bagdasaryan, Andreas Veit, Yiqing Hua, Deborah Estrin, and Vitaly
  Shmatikov.
\newblock How {To} {Backdoor} {Federated} {Learning}.
\newblock In {\em Proceedings of the {Twenty} {Third} {International}
  {Conference} on {Artificial} {Intelligence} and {Statistics}}, pages
  2938--2948. PMLR, June 2020.
\newblock ISSN: 2640-3498.

\bibitem{baruch_little_2019}
Gilad Baruch, Moran Baruch, and Yoav Goldberg.
\newblock A {Little} {Is} {Enough}: {Circumventing} {Defenses} {For}
  {Distributed} {Learning}.
\newblock In {\em Advances in {Neural} {Information} {Processing} {Systems}},
  volume~32. Curran Associates, Inc., 2019.

\bibitem{bhagoji_analyzing_2019}
Arjun~Nitin Bhagoji, Supriyo Chakraborty, Prateek Mittal, and Seraphin Calo.
\newblock Analyzing {Federated} {Learning} through an {Adversarial} {Lens}.
\newblock In {\em Proceedings of the 36th {International} {Conference} on
  {Machine} {Learning}}, pages 634--643. PMLR, May 2019.
\newblock ISSN: 2640-3498.

\bibitem{blanchard_machine_2017}
Peva Blanchard, El~Mahdi El~Mhamdi, Rachid Guerraoui, and Julien Stainer.
\newblock Machine {Learning} with {Adversaries}: {Byzantine} {Tolerant}
  {Gradient} {Descent}.
\newblock In {\em Advances in {Neural} {Information} {Processing} {Systems}},
  volume~30. Curran Associates, Inc., 2017.

\bibitem{boenisch_when_2021}
Franziska Boenisch, Adam Dziedzic, Roei Schuster, Ali~Shahin Shamsabadi, Ilia
  Shumailov, and Nicolas Papernot.
\newblock When the {Curious} {Abandon} {Honesty}: {Federated} {Learning} {Is}
  {Not} {Private}, December 2021.
\newblock arXiv:2112.02918 [cs].

\bibitem{BonawitzMPC2017}
Keith Bonawitz, Vladimir Ivanov, Ben Kreuter, Antonio Marcedone, H.~Brendan
  McMahan, Sarvar Patel, Daniel Ramage, Aaron Segal, and Karn Seth.
\newblock Practical secure aggregation for privacy-preserving machine learning.
\newblock In {\em Proceedings of the 2017 ACM SIGSAC Conference on Computer and
  Communications Security}, CCS '17, page 1175–1191, New York, NY, USA, 2017.
  Association for Computing Machinery.

\bibitem{cao_fltrust_2022}
Xiaoyu Cao, Minghong Fang, Jia Liu, and Neil~Zhenqiang Gong.
\newblock {FLTrust}: {Byzantine}-robust {Federated} {Learning} via {Trust}
  {Bootstrapping}, April 2022.
\newblock arXiv:2012.13995 [cs].

\bibitem{fang_local_2020}
Minghong Fang, Xiaoyu Cao, Jinyuan Jia, and Neil Gong.
\newblock Local {Model} {Poisoning} {Attacks} to \{{Byzantine}-{Robust}\}
  {Federated} {Learning}.
\newblock pages 1605--1622, 2020.

\bibitem{fowl_robbing_2022}
Liam Fowl, Jonas Geiping, Wojtek Czaja, Micah Goldblum, and Tom Goldstein.
\newblock Robbing the {Fed}: {Directly} {Obtaining} {Private} {Data} in
  {Federated} {Learning} with {Modified} {Models}, March 2022.
\newblock arXiv:2110.13057 [cs].

\bibitem{fredrikson_model_2015}
Matt Fredrikson, Somesh Jha, and Thomas Ristenpart.
\newblock Model {Inversion} {Attacks} that {Exploit} {Confidence} {Information}
  and {Basic} {Countermeasures}.
\newblock In {\em Proceedings of the 22nd {ACM} {SIGSAC} {Conference} on
  {Computer} and {Communications} {Security}}, {CCS} '15, pages 1322--1333, New
  York, NY, USA, October 2015. Association for Computing Machinery.

\bibitem{geiping_inverting_2020}
Jonas Geiping, Hartmut Bauermeister, Hannah Dröge, and Michael Moeller.
\newblock Inverting {Gradients} - {How} easy is it to break privacy in
  federated learning?
\newblock In {\em Advances in {Neural} {Information} {Processing} {Systems}},
  volume~33, pages 16937--16947. Curran Associates, Inc., 2020.

\bibitem{griffin2007caltech}
Gregory Griffin, Alex Holub, and Pietro Perona.
\newblock Caltech-256 object category dataset.
\newblock 2007.

\bibitem{he2016deep}
Kaiming He, Xiangyu Zhang, Shaoqing Ren, and Jian Sun.
\newblock Deep residual learning for image recognition.
\newblock In {\em Proceedings of the IEEE conference on computer vision and
  pattern recognition}, pages 770--778, 2016.

\bibitem{jeon_gradient_2021}
Jinwoo Jeon, jaechang Kim, Kangwook Lee, Sewoong Oh, and Jungseul Ok.
\newblock Gradient {Inversion} with {Generative} {Image} {Prior}.
\newblock In {\em Advances in {Neural} {Information} {Processing} {Systems}},
  volume~34, pages 29898--29908. Curran Associates, Inc., 2021.

\bibitem{kahla_label-only_2022}
Mostafa Kahla, Si~Chen, Hoang~Anh Just, and Ruoxi Jia.
\newblock Label-{Only} {Model} {Inversion} {Attacks} via {Boundary}
  {Repulsion}.
\newblock 2022.

\bibitem{kandpalDeduplicating22}
Nikhil Kandpal, Eric Wallace, and Colin Raffel.
\newblock Deduplicating training data mitigates privacy risks in language
  models.
\newblock In Kamalika Chaudhuri, Stefanie Jegelka, Le~Song, Csaba Szepesvari,
  Gang Niu, and Sivan Sabato, editors, {\em Proceedings of the 39th
  International Conference on Machine Learning}, volume 162 of {\em Proceedings
  of Machine Learning Research}, pages 10697--10707. PMLR, 17--23 Jul 2022.

\bibitem{karras_analyzing_2020}
Tero Karras, Samuli Laine, Miika Aittala, Janne Hellsten, Jaakko Lehtinen, and
  Timo Aila.
\newblock Analyzing and {Improving} the {Image} {Quality} of {StyleGAN}.
\newblock pages 8110--8119, 2020.

\bibitem{konecny_federated_2016}
Jakub Konečný, H.~Brendan McMahan, Daniel Ramage, and Peter Richtárik.
\newblock Federated {Optimization}: {Distributed} {Machine} {Learning} for
  {On}-{Device} {Intelligence}, October 2016.
\newblock arXiv:1610.02527 [cs].

\bibitem{krizhevsky2009learning}
Alex Krizhevsky, Geoffrey Hinton, et~al.
\newblock Learning multiple layers of features from tiny images.
\newblock 2009.

\bibitem{mcmahan_communication-efficient_2017}
Brendan McMahan, Eider Moore, Daniel Ramage, Seth Hampson, and Blaise Aguera~y
  Arcas.
\newblock Communication-{Efficient} {Learning} of {Deep} {Networks} from
  {Decentralized} {Data}.
\newblock In {\em Proceedings of the 20th {International} {Conference} on
  {Artificial} {Intelligence} and {Statistics}}, pages 1273--1282. PMLR, April
  2017.
\newblock ISSN: 2640-3498.

\bibitem{mhamdi_hidden_2018}
El~Mahdi~El Mhamdi, Rachid Guerraoui, and Sébastien Rouault.
\newblock The {Hidden} {Vulnerability} of {Distributed} {Learning} in
  {Byzantium}.
\newblock In {\em Proceedings of the 35th {International} {Conference} on
  {Machine} {Learning}}, pages 3521--3530. PMLR, July 2018.
\newblock ISSN: 2640-3498.

\bibitem{tiny-imagenet}
Mohammed~Ali mnmoustafa.
\newblock Tiny imagenet, 2017.

\bibitem{munoz-gonzalez_towards_2017}
Luis Muñoz-González, Battista Biggio, Ambra Demontis, Andrea Paudice, Vasin
  Wongrassamee, Emil~C. Lupu, and Fabio Roli.
\newblock Towards {Poisoning} of {Deep} {Learning} {Algorithms} with
  {Back}-gradient {Optimization}.
\newblock In {\em Proceedings of the 10th {ACM} {Workshop} on {Artificial}
  {Intelligence} and {Security}}, {AISec} '17, pages 27--38, New York, NY, USA,
  November 2017. Association for Computing Machinery.

\bibitem{shejwalkar_manipulating_2021}
Virat Shejwalkar and Amir Houmansadr.
\newblock Manipulating the {Byzantine}: {Optimizing} {Model} {Poisoning}
  {Attacks} and {Defenses} for {Federated} {Learning}.
\newblock In {\em Proceedings 2021 {Network} and {Distributed} {System}
  {Security} {Symposium}}, Virtual, 2021. Internet Society.

\bibitem{shen2022better}
Liyue Shen, Yanjun Zhang, Jingwei Wang, and Guangdong Bai.
\newblock Better together: Attaining the triad of byzantine-robust federated
  learning via local update amplification.
\newblock In {\em Proceedings of the 38th Annual Computer Security Applications
  Conference}, pages 201--213, 2022.

\bibitem{shokri_privacy-preserving_2015}
Reza Shokri and Vitaly Shmatikov.
\newblock Privacy-{Preserving} {Deep} {Learning}.
\newblock In {\em Proceedings of the 22nd {ACM} {SIGSAC} {Conference} on
  {Computer} and {Communications} {Security}}, {CCS} '15, pages 1310--1321, New
  York, NY, USA, October 2015. Association for Computing Machinery.

\bibitem{struppek_plug_2022}
Lukas Struppek, Dominik Hintersdorf, Antonio De~Almeida Correia, Antonia Adler,
  and Kristian Kersting.
\newblock Plug \& {Play} {Attacks}: {Towards} {Robust} and {Flexible} {Model}
  {Inversion} {Attacks}, June 2022.

\bibitem{usynin_beyond_2022}
Dmitrii Usynin, Daniel Rueckert, and Georgios Kaissis.
\newblock Beyond {Gradients}: {Exploiting} {Adversarial} {Priors} in {Model}
  {Inversion} {Attacks}, March 2022.
\newblock arXiv:2203.00481 [cs].

\bibitem{wang_attack_2020}
Hongyi Wang, Kartik Sreenivasan, Shashank Rajput, Harit Vishwakarma, Saurabh
  Agarwal, Jy-yong Sohn, Kangwook Lee, and Dimitris Papailiopoulos.
\newblock Attack of the {Tails}: {Yes}, {You} {Really} {Can} {Backdoor}
  {Federated} {Learning}.
\newblock In {\em Advances in {Neural} {Information} {Processing} {Systems}},
  volume~33, pages 16070--16084. Curran Associates, Inc., 2020.

\bibitem{wang2004image}
Zhou Wang, Alan~C Bovik, Hamid~R Sheikh, and Eero~P Simoncelli.
\newblock Image quality assessment: from error visibility to structural
  similarity.
\newblock {\em IEEE transactions on image processing}, 13(4):600--612, 2004.

\bibitem{wen_fishing_2022}
Yuxin Wen, Jonas Geiping, Liam Fowl, Micah Goldblum, and Tom Goldstein.
\newblock Fishing for {User} {Data} in {Large}-{Batch} {Federated} {Learning}
  via {Gradient} {Magnification}, June 2022.
\newblock arXiv:2202.00580 [cs].

\bibitem{xie_generalized_2018}
Cong Xie, Oluwasanmi Koyejo, and Indranil Gupta.
\newblock Generalized {Byzantine}-tolerant {SGD}, March 2018.
\newblock arXiv:1802.10116 [cs, stat].

\bibitem{ye_model_2022}
Dayong Ye, Huiqiang Chen, Shuai Zhou, Tianqing Zhu, Wanlei Zhou, and Shouling
  Ji.
\newblock Model {Inversion} {Attack} against {Transfer} {Learning}: {Inverting}
  a {Model} without {Accessing} {It}, March 2022.
\newblock arXiv:2203.06570 [cs].

\bibitem{yin_byzantine-robust_2018}
Dong Yin, Yudong Chen, Ramchandran Kannan, and Peter Bartlett.
\newblock Byzantine-{Robust} {Distributed} {Learning}: {Towards} {Optimal}
  {Statistical} {Rates}.
\newblock In {\em Proceedings of the 35th {International} {Conference} on
  {Machine} {Learning}}, pages 5650--5659. PMLR, July 2018.
\newblock ISSN: 2640-3498.

\bibitem{yin_see_2021}
Hongxu Yin, Arun Mallya, Arash Vahdat, Jose~M. Alvarez, Jan Kautz, and Pavlo
  Molchanov.
\newblock See {Through} {Gradients}: {Image} {Batch} {Recovery} via
  {GradInversion}.
\newblock pages 16337--16346, 2021.

\bibitem{yin_dreaming_2020}
Hongxu Yin, Pavlo Molchanov, Jose~M. Alvarez, Zhizhong Li, Arun Mallya, Derek
  Hoiem, Niraj~K. Jha, and Jan Kautz.
\newblock Dreaming to {Distill}: {Data}-{Free} {Knowledge} {Transfer} via
  {DeepInversion}.
\newblock pages 8715--8724, 2020.

\bibitem{zhang2020batchcrypt}
Chengliang Zhang, Suyi Li, Junzhe Xia, Wei Wang, Feng Yan, and Yang Liu.
\newblock Batchcrypt: {E}fficient homomorphic encryption for cross-silo
  federated learning.
\newblock In {\em Proceedings of the 2020 USENIX Annual Technical Conference
  (USENIX ATC 2020)}, 2020.

\bibitem{zhang2018unreasonable}
Richard Zhang, Phillip Isola, Alexei~A Efros, Eli Shechtman, and Oliver Wang.
\newblock The unreasonable effectiveness of deep features as a perceptual
  metric.
\newblock In {\em Proceedings of the IEEE conference on computer vision and
  pattern recognition}, pages 586--595, 2018.

\bibitem{zhang2022survey}
Rui Zhang, Song Guo, Junxiao Wang, Xin Xie, and Dacheng Tao.
\newblock A survey on gradient inversion: Attacks, defenses and future
  directions.
\newblock {\em arXiv preprint arXiv:2206.07284}, 2022.

\bibitem{zhao_idlg_2020}
Bo~Zhao, Konda~Reddy Mopuri, and Hakan Bilen.
\newblock {iDLG}: {Improved} {Deep} {Leakage} from {Gradients}, January 2020.
\newblock arXiv:2001.02610 [cs, stat].

\bibitem{zhu_r-gap_2022}
Junyi Zhu and Matthew~B. Blaschko.
\newblock R-{GAP}: {Recursive} {Gradient} {Attack} on {Privacy}.
\newblock February 2022.

\bibitem{zhu_deep_2019}
Ligeng Zhu, Zhijian Liu, and Song Han.
\newblock Deep {Leakage} from {Gradients}.
\newblock In {\em Advances in {Neural} {Information} {Processing} {Systems}},
  volume~32. Curran Associates, Inc., 2019.

\end{thebibliography}

\appendix

\renewcommand{\thesection}{Appendix~\Alph{section}.}
\renewcommand{\thesubsection}{\Alph{section}.\arabic{subsection}}
\setcounter{table}{0}   
\setcounter{figure}{0}
\renewcommand{\thetable}{\Alph{section}\arabic{table}}
\renewcommand{\thefigure}{\Alph{section}\arabic{figure}}

\section{Symbol Tables}~\label{appendix:symbol_tables}

\begin{table}[H]
\centering
\caption{Basic Symbols.}
\small
\begin{tblr}{
  cells = {c},
  hline{1,2,12} = {-}{}
}
Symbol & Description \\
$t$ & The current round in FL. \\
$N$ & The total number of clients. \\
$D_i$ & The $i$'th client's private dataset. \\
$\Phi_G^t$ & The parameters of the global model at $t$'th round. \\
$\delta_i^t$ & The submitted update from the $i$'th client at $t$'th round. \\
$|B|$ & The batch size. \\
$(x_i,y_i)$ & The $i$'th image and its label. \\
$(x_{tar},y_{tar})$ & The targeted class image and label pair input. \\
$\delta_{tar}$ & Gradient of the targeted class samples.\\
$\delta_{batch}$ & Gradient of the batch.\\
\end{tblr}
\end{table}

\begin{table}[H]
\centering
\caption{Symbols in \attack.}
\small
\begin{tblr}{
  cells = {c,m},
  hline{1,2,12} = {-}{}
}
Symbol & Description \\
$\mathcal{K}$ & The knowledge that the adversary has. \\
$D_{loc}$ & The local dataset belonging to the malicious client. \\
$\alpha$ & The learning rate. \\
$\lambda$ & {The coefficients of regularization items \\ in the inversion formula.} \\
$\Phi_{mal}$ & The locally trained malicious model. \\
$\delta_{poi}$ & {The poisoning replacement vector between \\ the global model and the malicious model.} \\
$\delta_{agr}$ & {The global aggregation gradient after \\ filtering malicious updates by AGR.}\\
$\delta_{m}$ & The submitted malicious update of malicious clients.\\
$\overline{\delta}_{ben}$ & The benign gradient from malicious clients. \\
$\hat{\delta}_{tar}$ & The gradient of dummy targeted samples. \\
\end{tblr}

\end{table}

\section{Technical Details of Experimental Setting}~\label{appendix:details}
\vspace{-0.2cm}
\subsection{Dataset}
We use Cifar100~\cite{krizhevsky2009learning}, TinyImageNet~\cite{tiny-imagenet}, and CalTech256~\cite{griffin2007caltech} as benchmark datasets and the introductions of them as follows:  
\begin{itemize}
    \item \textbf{Cifar100}~\cite{krizhevsky2009learning} is a 100-class classification task dataset with 60,000 RGB images. All images are of size 32 x 32. There are 500 training images and 100 testing images per class. 

    \item \textbf{TinyImageNet}~\cite{tiny-imagenet} is a subset of ImageNet. It contains 100,000 images from 200 classes. Each image is a 64 x 64 color image. 

    \item \textbf{CalTech256}~\cite{griffin2007caltech} is an object recognition dataset containing 30,607 real-world images, of different sizes, spanning 257 classes. We scaled the images to 112 x 112 and selected 256 classes in our task. 
\end{itemize}

\subsection{Hyper-parameter settings} 
We detail the hyper-parameter settings in the three steps respectively.

\paragraph{Hyper-parameters in crafting the local malicious model}
We set $\alpha_1 = 1\times10^{-2}$, $\alpha_2 = 1\times10^{-5}$ for Algorithm~\ref{alg:MR} (in the setting of full knowledge) for all datasets. 
We set $\alpha_1 = 5\times10^{-3}$ for all datasets for Algorithm~\ref{alg:MP} (in the setting of semi knowledge) for the poisoning process, and set a small learning rate for the repair step i.e., $\alpha_2 = 1\times10^{-4}, 1\times10^{-4}, 1\times10^{-5}$ for \textit{Cifar100}, \textit{TinyImageNet}, and \textit{CalTech256}, respectively, to avoid overfitting on auxiliary datasets. 
For the hyper-parameters in Algorithm~\ref{alg:MIP} (in the setting of no knowledge), we use $\lambda_0 = 1\times10^{-1},1\times10^{-2},1\times10^{-2}$; $\lambda_1 = 1\times10^{-5},1\times10^{-6},1\times10^{-3}$; $\lambda_2 = 1\times10^{-3},1\times10^{-2},1\times10^{-5}$; $\lambda_3 = 1,1\times10^{-1},1\times10^{-1}$ for \textit{Cifar100}, \textit{TinyImageNet}, and \textit{CalTech256}, respectively. For the learning rates, we set $\alpha_1 = 1\times10^{-3}$, $\alpha_2 = 1\times10^{-5}$ for all datasets.


\paragraph{Hyper-parameters in poisoning the global model}
Basically, the hyperparameters setting of poisoning varies dramatically for different knowledge levels, AGRs, and datasets. For Algorithm~\ref{alg:OptMalUpdate}, we set learning rates $\alpha_1$ and $\alpha_2$ a very small value (e.g., $1\times10^{-5}$ or $1\times10^{-6}$) for optimizing $\gamma$ and $\beta$. Then initialized values $\gamma_0$ and $\beta_0$ represent the initialized magnitude of maliciousness and concealment in the malicious update. In practice, the adversary needs to observe the model distance between the global model and the malicious model in each round to adjust $\gamma_0$ and $\beta_0$. If the model distance increases, which means the malicious update submitted at the previous round has been filtered by AGR, the adversary should set a smaller $\gamma_0$ and a larger $\beta_0$. But if the model distance decreases very slowly, the adversary can try to increase $\gamma_0$ and speed up poisoning. Here, we just give a group hyperparameter for each AGR as an example. We use $\gamma_0 = 10$, $\beta_0 = 1$; $\gamma_0 = 1\times10^{-2}$, $\beta_0 = 1$; $\gamma_0 = 10$, $\beta_0 = 1\times10^{-2}$; $\gamma_0 = 10$, $\beta_0 = 1$ for \textit{Multi-Krum}, \textit{Bulyan}, \textit{AFA} and \textit{Fang} respectively.

\paragraph{Hyper-parameters in inverting the targeted gradients}
We set $\lambda_0 = 10$; $\lambda_1 = 1\times10^{-2}$; $\lambda_2 = 1\times10^{-6}$; $\lambda_3 = 1.0$ for Equation~(\ref{equ:inversion}). 

\section{Effect of the Targeted Sample's Distribution}~\label{appendix:targeted_distribution}
\setcounter{table}{0}   
\setcounter{figure}{0}
We also consider the case in which more than one targeted samples appear in one aggregation. 
To this end, we study the effect of the targeted sample distribution and investigate two typical settings: the first one is that all targeted samples belong to a single client (single-client distribution), and the second one is that the targeted samples are distributed uniformly at random among multiple clients (multi-client distribution). 
We fix the number of clients as 40, the batch size as 32, and the number of malicious clients as 8. The FL is equipped with Multi-krum~\cite{blanchard_machine_2017} and we assume the adversary has full knowledge. 

\begin{figure}[!htp]
	\centering
	\subfloat[Multiple-client setting]{\includegraphics[width=\linewidth]{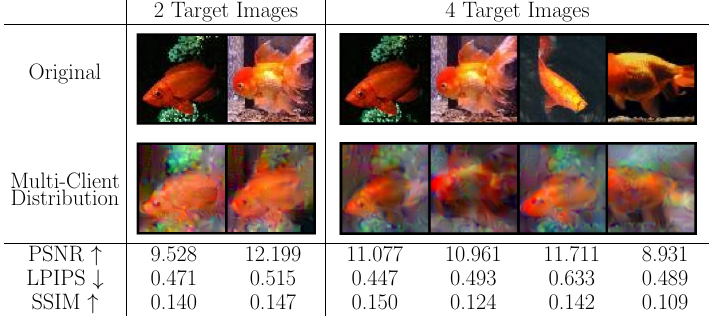}%
		}
    \\
	\subfloat[Single-client setting]{\includegraphics[width=\linewidth]{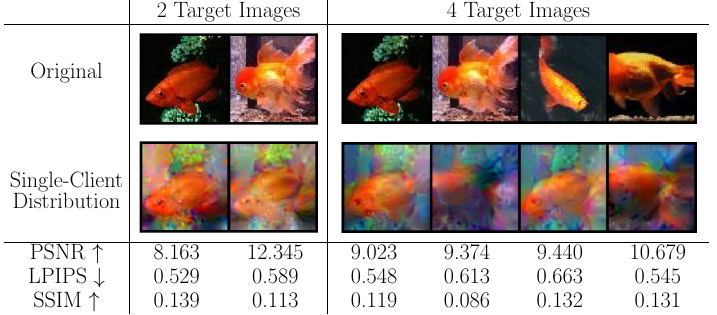}%
	}
    \caption{\attack's performance of restoring multiple targeted samples on TinyImageNet.}
    \label{fig:MultiInversion}
\end{figure}

Figure~\ref{fig:MultiInversion} illustrates the results. In general, our \attack succeeds in restoring multiple targeted samples from one aggregation. 
However, we observe some transfers of semantic features among the restored samples of which their corresponding original samples are similar because their gradients are averaged during the aggregation. 

We also find that such transfers of semantic features are more likely to appear in single-client distribution. 
This is due to the fact that targeted samples are aggregated locally, which happens prior to the global aggregation. When all targeted samples are grouped together in a single batch and processed by a single client, the batch data will go through the same set of batch normalization layers. As a result, the batch normalization layers facilitate the transfer of similar image features by reducing internal covariate shifts and stabilizing the feature distribution.



\section{Effect of the Number of Clients}~\label{appendix:number_client}
Here, we present the result of \attack performance against the number of clients when fixing the batch size 128. We can find that the tendency is similar to our statement in Section~\ref{sec:ablation}.
\setcounter{table}{0}   
\setcounter{figure}{0}
\begin{figure}[!htp]
	\centering
	\subfloat[PSNR]{\includegraphics[width=2in]{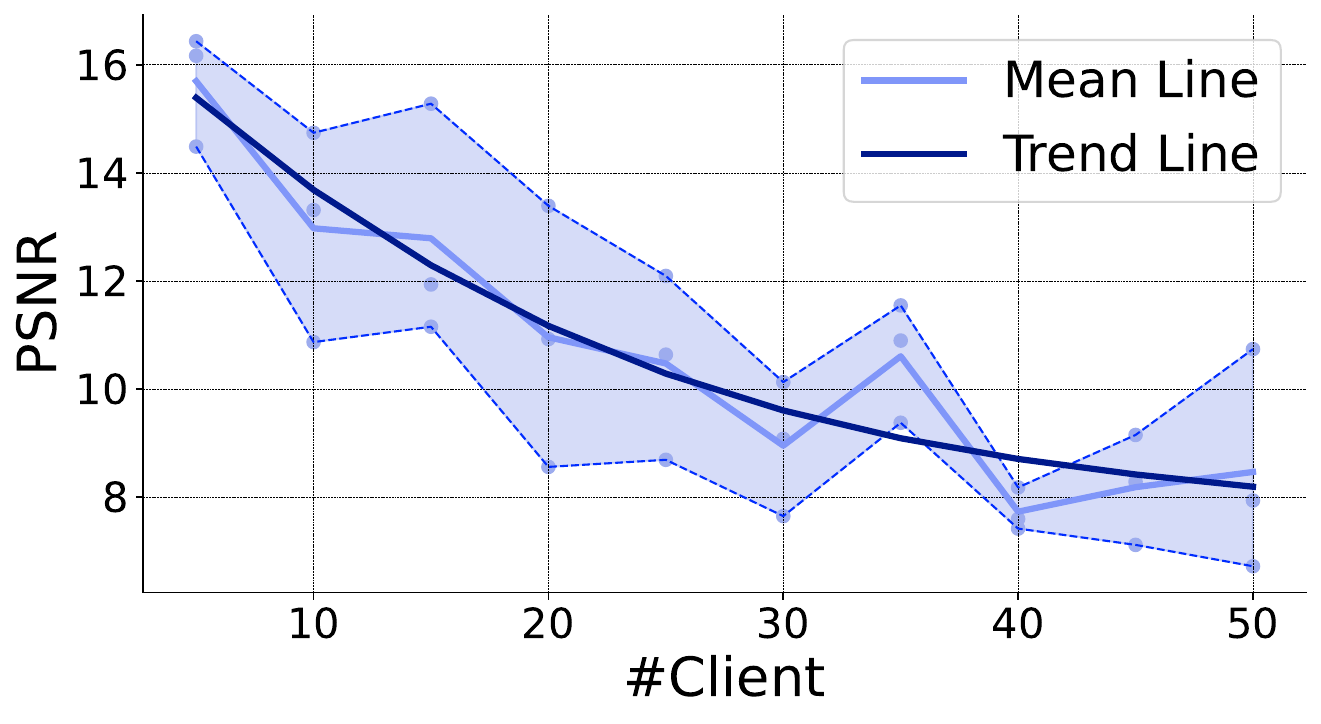}%
		\label{Fig_BatchEffectSubPSNR}} \\
	\subfloat[LPIPS]{\includegraphics[width=2in]{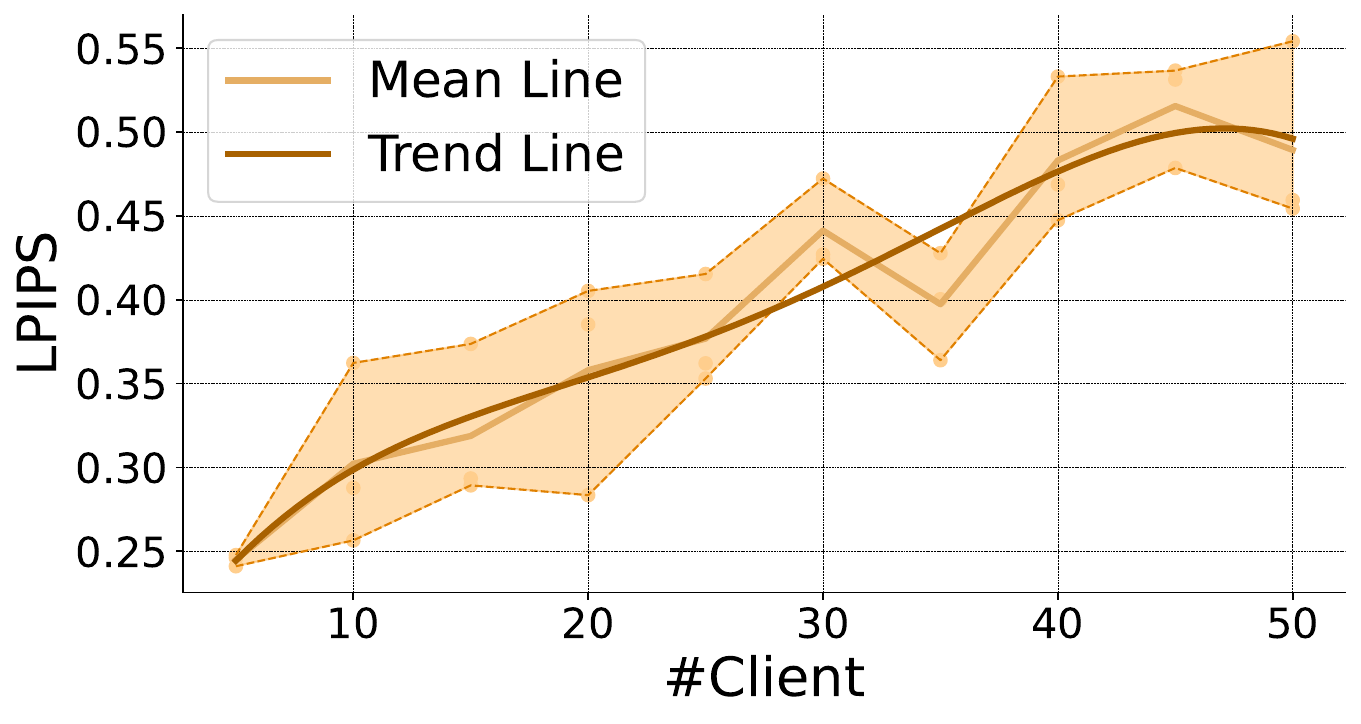}%
		\label{Fig_BatchEffectSubLPIPS}} \\
	\subfloat[SSIM]{\includegraphics[width=2in]{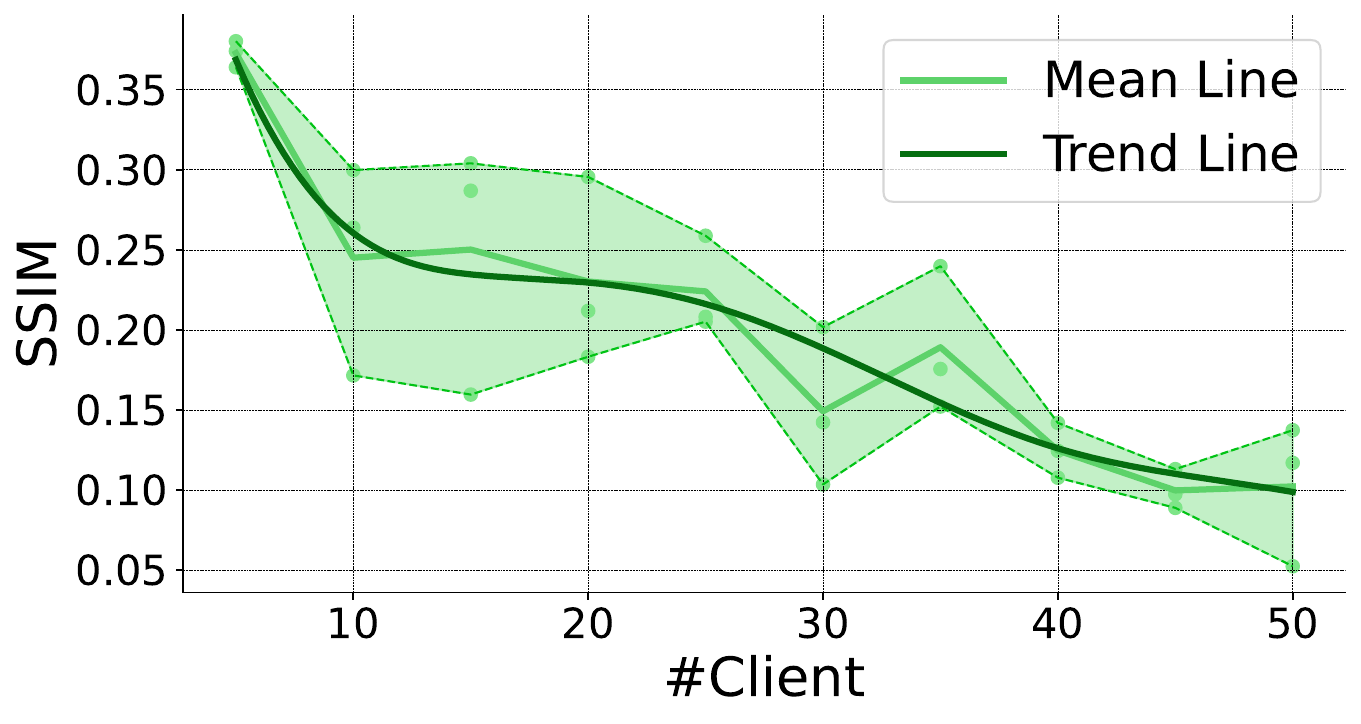}%
		\label{Fig_BatchEffectSubSSIM}} 
	\caption{PSNR, LPIPS, and SSIM of \attack against the client number with batch size 128 on TinyImageNet.}
	\label{fig:BatchEffectChart2}
\end{figure}

\end{document}